%% file: main.tex
\pdfoutput=1
\documentclass[12pt,a4paper]{article}

\usepackage{ifthen} 
\newboolean{pdflatex}
\setboolean{pdflatex}{true} 

\newboolean{articletitles}
\setboolean{articletitles}{true} 

\newboolean{uprightparticles}
\setboolean{uprightparticles}{false} 

\def\paperauthors{LHCb collaboration} 
\def\paperasciititle{Absolute luminosity calibration of the LHCb detector in in pp collisions at sqrt{s} = 13 TeV with beam-gas imaging} 
\def\papertitle{Absolute luminosity calibration of the \lhcb detector in \proton\proton collisions at \mbox{\sqs=13\tev} with beam-gas imaging} 
\def\paperkeywords{{High Energy Physics}, {LHCb}} 
\def\papercopyright{\the\year\ CERN for the benefit of the LHCb collaboration} 
\def\paperlicence{CC BY 4.0 licence}
\def\paperlicenceurl{https://creativecommons.org/licenses/by/4.0/}

\newif\ifEnableSectionTOCLinks
\EnableSectionTOCLinksfalse 

\input{LHCb/preamble}
\usepackage{longtable} 

\input{local-symbols-def} 

\usepackage{subcaption}

\begin{document}

\renewcommand{\thefootnote}{\fnsymbol{footnote}}
\setcounter{footnote}{1}

\input{title-LHCb-PAPER}


\renewcommand{\thefootnote}{\arabic{footnote}}
\setcounter{footnote}{0}


\cleardoublepage


\pagestyle{plain} 
\setcounter{page}{1}
\pagenumbering{arabic}


\input{body}

\input{LHCb/acknowledgements}



\addcontentsline{toc}{section}{References}
\bibliographystyle{LHCb/LHCb}
\bibliography{main,LHCb/standard,LHCb/LHCb-PAPER,LHCb/LHCb-CONF,LHCb/LHCb-DP,LHCb/LHCb-TDR}

\newpage
\input{Authorship_LHCb-PAPER-2026-025}

\end{document}

%% file: LHCb/preamble.tex
\usepackage[top=1in, bottom=1.25in, left=1in, right=1in]{geometry}

\usepackage{microtype}
\usepackage{lineno}  
\usepackage{xspace} 
\usepackage{caption} 

\usepackage{graphicx}  
\usepackage{color}
\usepackage{colortbl}
\graphicspath{{./figs/}} 

\usepackage{amsmath} 
\usepackage{amssymb}
\usepackage{amsfonts}
\usepackage{upgreek} 

\usepackage[normalem]{ulem} 

\newcommand*\patchAmsMathEnvironmentForLineno[1]{%
\expandafter\let\csname old#1\expandafter\endcsname\csname #1\endcsname
\expandafter\let\csname oldend#1\expandafter\endcsname\csname
end#1\endcsname
 \renewenvironment{#1}%
   {\linenomath\csname old#1\endcsname}%
   {\csname oldend#1\endcsname\endlinenomath}%
}
\newcommand*\patchBothAmsMathEnvironmentsForLineno[1]{%
  \patchAmsMathEnvironmentForLineno{#1}%
  \patchAmsMathEnvironmentForLineno{#1*}%
}
\AtBeginDocument{%
\patchBothAmsMathEnvironmentsForLineno{equation}%
\patchBothAmsMathEnvironmentsForLineno{align}%
\patchBothAmsMathEnvironmentsForLineno{flalign}%
\patchBothAmsMathEnvironmentsForLineno{alignat}%
\patchBothAmsMathEnvironmentsForLineno{gather}%
\patchBothAmsMathEnvironmentsForLineno{multline}%
\patchBothAmsMathEnvironmentsForLineno{eqnarray}%
}

\usepackage[pdftex,
            pdfauthor={\paperauthors},
            pdftitle={\paperasciititle},
            pdfkeywords={\paperkeywords}]{hyperref}
\usepackage{hyperxmp}
\hypersetup{
    pdfcopyright={Copyright (C) \papercopyright},
    pdflicenseurl={\paperlicenceurl}
}

\usepackage[colorinlistoftodos,textsize=scriptsize]{todonotes}

\usepackage[bottom,flushmargin,hang,multiple]{footmisc}

\usepackage[all]{hypcap} 

\input{LHCb/lhcb-symbols-def} 

\hypersetup{
  colorlinks   = true, 
  urlcolor     = blue, 
  linkcolor    = blue, 
  citecolor    = red   
}

\ifEnableSectionTOCLinks
    \usepackage[explicit]{titlesec} 
    
    \let\oldcontentsline\contentsline
    \renewcommand

    \titleformat{\section}{\normalfont\Large\bf}{\hyperlink{tocsection.\thesection}{{\thesection} \parbox[t]{\dimexpr\textwidth-1pc}{#1}}}{1pc}{}

    \titleformat{\subsection}{\normalfont\bf}{\hyperlink{tocsubsection.\thesubsection}{{\thesubsection} \parbox[t]{\dimexpr\textwidth-1pc}{#1}}}{1pc}{}

    \titleformat{name=\section,numberless}[display]{}{}{0pt}{\normalfont\Huge\bfseries #1}
\fi

\usepackage{cite} 
\usepackage{LHCb/mciteplus}

%% file: LHCb/lhcb-symbols-def.tex
\usepackage{xspace} 
\usepackage{upgreek}

\def\lhcb   {\mbox{LHCb}\xspace}

\def\lhc    {\mbox{LHC}\xspace}

\def\velo   {VELO\xspace}

\def\MagUp {\mbox{\em Mag\kern -0.05em Up}\xspace}

\ifthenelse{\boolean{uprightparticles}}%
{

 \def\PDelta      {\ensuremath{\Delta}\xspace}                 
 \def\PXi         {\ensuremath{\Xi}\xspace}                 
 \def\PLambda     {\ensuremath{\Lambda}\xspace}                 
 \def\PSigma      {\ensuremath{\Sigma}\xspace}                 
 \def\POmega      {\ensuremath{\Omega}\xspace}                 
 \def\PUpsilon    {\ensuremath{\Upsilon}\xspace}
 \let\oldPi\Pi
 \def\PPi         {\ensuremath{\oldPi}\xspace}

 \def\PB      {\ensuremath{\mathrm{B}}\xspace}                 
 \def\PD      {\ensuremath{\mathrm{D}}\xspace}                 

 \def\PK      {\ensuremath{\mathrm{K}}\xspace}                 
 \def\Pb      {\ensuremath{\mathrm{b}}\xspace}                 
 \def\Pc      {\ensuremath{\mathrm{c}}\xspace}

 \def\Pp      {\ensuremath{\mathrm{p}}\xspace}                 

 \def\Ps      {\ensuremath{\mathrm{s}}\xspace}

 \def\thebaroffset{0.0em}
}
{

 \mathchardef\PDelta="7101
 \mathchardef\PXi="7104
 \mathchardef\PLambda="7103
 \mathchardef\PSigma="7106
 \mathchardef\POmega="710A
 \mathchardef\PUpsilon="7107
 \mathchardef\PPi="7105
 \def\PB      {\ensuremath{B}\xspace}                 
 \def\PD      {\ensuremath{D}\xspace}                 

 \def\PK      {\ensuremath{K}\xspace}                 
 \def\Pb      {\ensuremath{b}\xspace}                 
 \def\Pc      {\ensuremath{c}\xspace}

 \def\Pp      {\ensuremath{p}\xspace}                 

 \def\Ps      {\ensuremath{s}\xspace}

 \def\thebaroffset{0.18em}
}
\newcommand{\offsetoverline}[2][\thebaroffset]{\kern #1\overline{\kern -#1 #2}}%

\makeatletter
\ifcase \@ptsize \relax
  \newcommand{\miniscule}{\@setfontsize\miniscule{4}{5}}
\or
  \newcommand{\miniscule}{\@setfontsize\miniscule{5}{6}}
\or
  \newcommand{\miniscule}{\@setfontsize\miniscule{5}{6}}
\fi
\makeatother

\DeclareRobustCommand{\optbar}[1]{\shortstack{{\miniscule (\rule[.5ex]{1.25em}{.18mm})}
  \\ [-.7ex] $#1$}}

\def\squark    {{\ensuremath{\Ps}}\xspace}

\def\cquark    {{\ensuremath{\Pc}}\xspace}

\def\bquark    {{\ensuremath{\Pb}}\xspace}

\def\KorKbar {\kern \thebaroffset\optbar{\kern -\thebaroffset \PK}{}\xspace}

\def\D       {{\ensuremath{\PD}}\xspace}

\def\DorDbar {\kern \thebaroffset\optbar{\kern -\thebaroffset \PD}\xspace}

\def\Dp      {{\ensuremath{\D^+}}\xspace}
\def\Dm      {{\ensuremath{\D^-}}\xspace}

\def\DpDm    {\ensuremath{\Dp {\kern -0.16em \Dm}}\xspace}

\def\B       {{\ensuremath{\PB}}\xspace}

\def\BorBbar {\kern \thebaroffset\optbar{\kern -\thebaroffset \PB}\xspace}

\def\Bd      {{\ensuremath{\B^0}}\xspace}

\def\BdorBdbar {\kern \thebaroffset\optbar{\kern -\thebaroffset \Bd}\xspace}

\def\Bs      {{\ensuremath{\B^0_\squark}}\xspace}

\def\BsorBsbar {\kern \thebaroffset\optbar{\kern -\thebaroffset \Bs}\xspace}

\def\Y#1S{\ensuremath{\PUpsilon{(#1S)}}\xspace}

\def\proton      {{\ensuremath{\Pp}}\xspace}

\def\LorLbar     {\kern \thebaroffset\optbar{\kern -\thebaroffset \PLambda}\xspace}

\def\AT#1     {\ensuremath{A_{\mathrm{T}}^{#1}}\xspace}           

\def\C#1      {\ensuremath{\mathcal{C}_{#1}}\xspace}                       
\def\Cp#1     {\ensuremath{\mathcal{C}_{#1}^{'}}\xspace}                    
\def\Ceff#1   {\ensuremath{\mathcal{C}_{#1}^{\mathrm{(eff)}}}\xspace}        
\def\Cpeff#1  {\ensuremath{\mathcal{C}_{#1}^{'\mathrm{(eff)}}}\xspace}       
\def\Ope#1    {\ensuremath{\mathcal{O}_{#1}}\xspace}                       
\def\Opep#1   {\ensuremath{\mathcal{O}_{#1}^{'}}\xspace}                    

\newcommand{\nospaceunit}[1]{\ensuremath{\text{#1}}}       
\newcommand{\aunit}[1]{\ensuremath{\text{\,#1}}}       

\newcommand{\tev}{\aunit{Te\kern -0.1em V}\xspace}
\newcommand{\gev}{\aunit{Ge\kern -0.1em V}\xspace}
\newcommand{\mev}{\aunit{Me\kern -0.1em V}\xspace}
\newcommand{\kev}{\aunit{ke\kern -0.1em V}\xspace}
\newcommand{\ev}{\aunit{e\kern -0.1em V}\xspace}
 
\newcommand{\mevc}{\ensuremath{\aunit{Me\kern -0.1em V\!/}c}\xspace}
\newcommand{\gevc}{\ensuremath{\aunit{Ge\kern -0.1em V\!/}c}\xspace}
\newcommand{\mevcc}{\ensuremath{\aunit{Me\kern -0.1em V\!/}c^2}\xspace}
\newcommand{\gevcc}{\ensuremath{\aunit{Ge\kern -0.1em V\!/}c^2}\xspace}

\def\m    {\aunit{m}\xspace}

\def\mm   {\aunit{mm}\xspace}

\def\mum  {\ensuremath{\,\upmu\nospaceunit{m}}\xspace}

\def\mbarn{\aunit{mb}\xspace}

\def\sec  {\ensuremath{\aunit{s}}\xspace}

\def\ns   {\ensuremath{\aunit{ns}}\xspace}

\def\mhz  {\ensuremath{\aunit{MHz}}\xspace}
\def\khz  {\ensuremath{\aunit{kHz}}\xspace}

\def\hr   {\aunit{hr}\xspace}

\newcommand{\stat}{\aunit{(stat)}\xspace}
\newcommand{\syst}{\aunit{(syst)}\xspace}

\def\gsim{{~\raise.15em\hbox{$>$}\kern-.85em
          \lower.35em\hbox{$\sim$}~}\xspace}
\def\lsim{{~\raise.15em\hbox{$<$}\kern-.85em
          \lower.35em\hbox{$\sim$}~}\xspace}

\def\sqs   {\ensuremath{\protect\sqrt{s}}\xspace}

\def\betastar {\ensuremath{\beta^*}}

\def\tell1  {TELL1\xspace}
\def\ukl1   {UKL1\xspace}

\newcommand{\ie}{\mbox{\itshape i.e.}\xspace}

\newcommand{\lhcborcid}[1]{\href{https://orcid.org/#1}{\hspace*{0.1em}\raisebox{-0.45ex}{\includegraphics[width=1em]{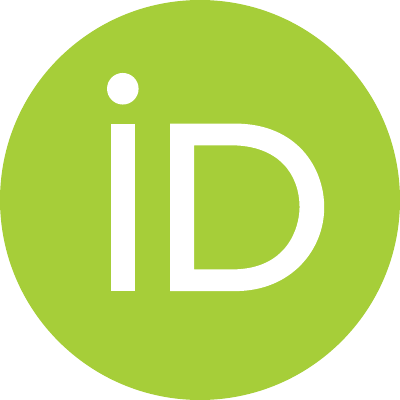}}}}


%% file: local-symbols-def.tex
\newcommand{\ntot}{{\ensuremath{n_{\mathrm{tot}}}}\xspace}

%% file: title-LHCb-PAPER.tex

\begin{titlepage}
\pagenumbering{roman}

\vspace*{-1.5cm}
\centerline{\large EUROPEAN ORGANIZATION FOR NUCLEAR RESEARCH (CERN)}
\vspace*{1.5cm}
\noindent
\begin{tabular*}{\linewidth}{lc@{\extracolsep{\fill}}r@{\extracolsep{0pt}}}
\ifthenelse{\boolean{pdflatex}}
{\vspace*{-1.5cm}\mbox{\!\!\!\includegraphics[width=.14\textwidth]{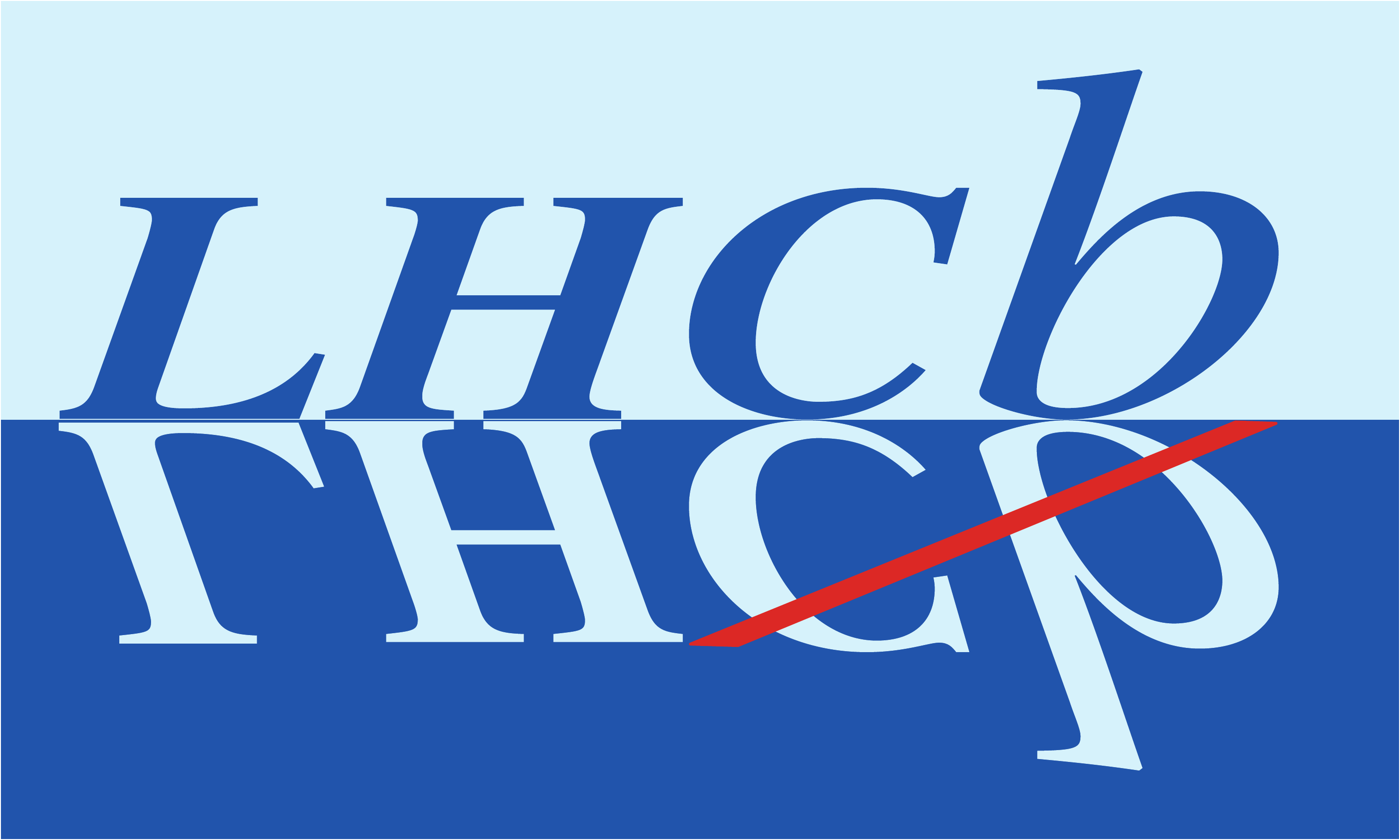}} & &}%
{\vspace*{-1.2cm}\mbox{\!\!\!\includegraphics[width=.12\textwidth]{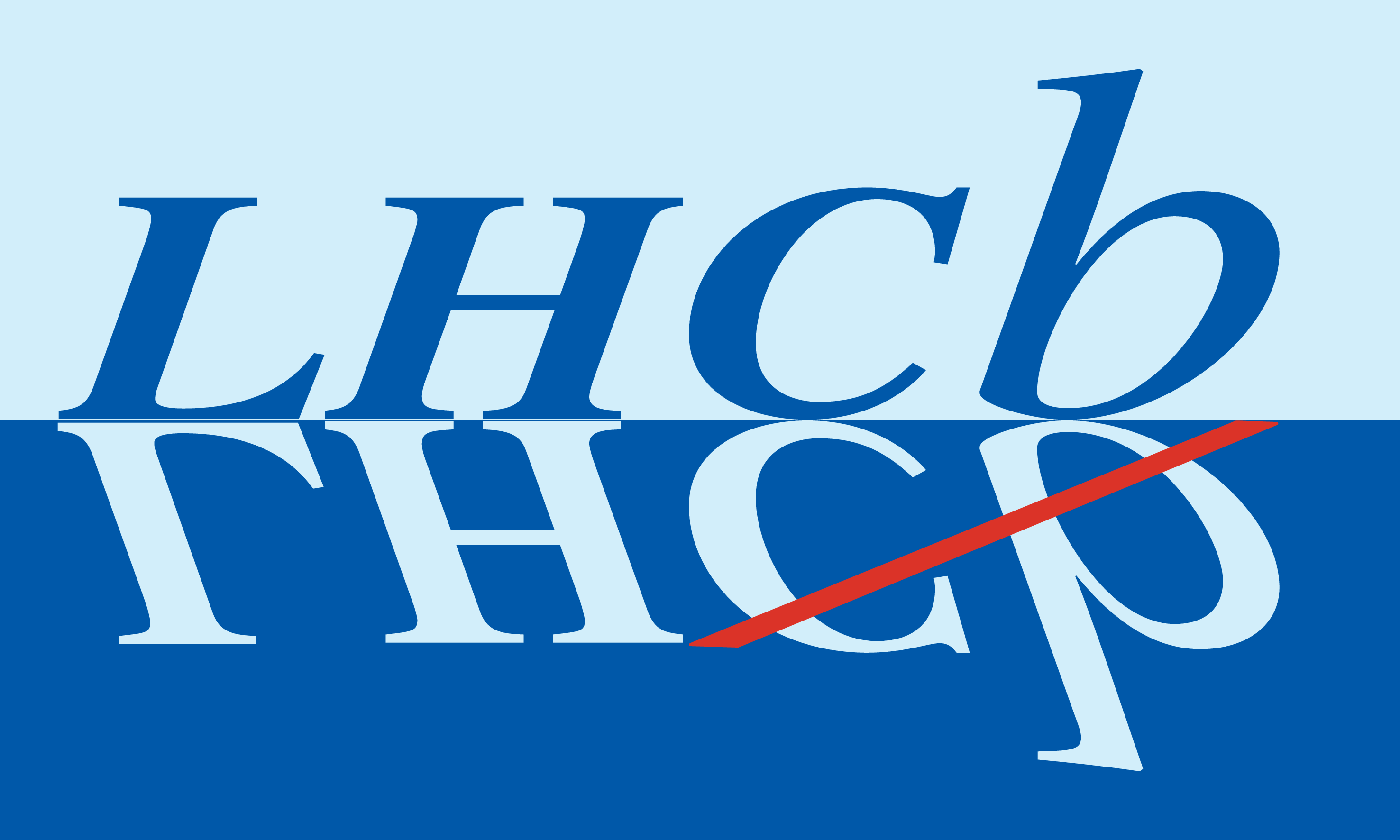}} & &}%
\\
 & & CERN-EP-2026-282 \\  
 & & LHCb-PAPER-2026-025 \\  
 & & November 17, 2026 \\ 
 & & \\
\end{tabular*}

\vspace*{4.0cm}

{\normalfont\bfseries\boldmath\huge
\begin{center}
  \papertitle 
\end{center}
}

\vspace*{2.0cm}

\begin{center}
\paperauthors\footnote{Authors are listed at the end of this paper.}
\end{center}

\vspace{\fill}

\begin{abstract}
  \noindent
An absolute calibration of the luminosity for proton-proton collisions at \mbox{$\sqrt{s} = 13$\tev} with the \lhcb detector in Run 2 of the \lhc is presented. The calibration cross-section of the Vertex counter, a statistic based on reconstructed primary vertices, is measured using the beam-gas imaging method to be \mbox{$\sigma_\mathrm{Vertex} = 57.61 \pm 0.06 \stat \pm 0.52 \syst \mbarn$}, corresponding to a relative precision of 0.91\%. This result provides around a factor of two improvement in the precision of the \lhcb luminosity calibration used in previous cross-section measurements with this dataset. The largest systematic uncertainties arise from the imperfect knowledge of the beam-gas vertex resolution, the observed fill-by-fill spread of the cross-section, and the choice of transverse beam-profile fit model.
  
\end{abstract}

\vspace*{2.0cm}

\begin{center}
  Submitted to JINST
\end{center}

\vspace{\fill}

{\footnotesize 
\centerline{\copyright~\papercopyright. \href{\paperlicenceurl}{\paperlicence}.}}
\vspace*{2mm}

\end{titlepage}


\newpage
\setcounter{page}{2}
\mbox{~}
%
%
%
%

%% file: body.tex
\section{Introduction}
\label{sec:Introduction}

The measurement of the absolute luminosity forms a crucial input for any absolute cross-section measurement. For a single colliding bunch pair in a collider the integrated luminosity, $\mathcal{L}$, can be defined as
\begin{equation}
\label{eq:lumi-rate}
    \mathcal{L} = n\frac{\mu}{\sigma},
\end{equation}
where $n$ is the number of times the bunches cross, and $\mu$ and $\sigma$ are the mean number of interactions per crossing and cross-section, respectively, for a given process. At the LHC, the absolute luminosity is typically calibrated using the van der Meer (vdM) approach due to the absence of a process with a cross-section that is both theoretically well-known and experimentally well-defined~\cite{vanderMeer:296752, LHCb-PAPER-2014-047, ALICE_vdm, ATLAS_vdm, CMS_vdm}. In addition, the \lhcb experiment has the unique capability of calibrating the luminosity by directly reconstructing the individual beam profiles using the beam-gas imaging (BGI) method~\cite{FerroLuzzi:2005em, LHCb-PAPER-2014-047, Barschel:2014iua, Coombs:2021jmj, McHugh:2024pnu}. 

The excellent primary collision vertex (PV) resolution of the \lhcb VErtex LOcator (VELO)~\cite{LHCb-TDR-005, LHCb-DP-2014-001} and increased beam-gas collision rate when injecting gas into the beam vacuum chamber with the System for Measuring Overlap with Gas (SMOG)~\cite{FerroLuzzi:2005em, Barschel:2014iua} allow a determination of the beam profiles by studying the distribution of beam-gas vertices. The integrated luminosity can be written in terms of the beam parameters as
\begin{equation}
\label{eq:lumi-beam}
    \mathcal{L} =  n N_1 N_2 K \mathcal{O},
\end{equation}
where $N_1$ and $N_2$ are the number of protons per bunch in each beam, $\rho_i(\vec{x}, t)$ is the profile density of the colliding bunch in beam $i$, and 
\begin{equation}
    \mathcal{O} \equiv \int \rho_1(\vec{x}, t)\, \rho_2(\vec{x}, t)\, d\vec{x}\, dt.
\end{equation}
The term $K \equiv \sqrt{(\vec{v}_1 - \vec{v}_2)^2 - (\vec{v}_1 \times \vec{v}_2)^2/c^2}$ is a kinematic factor that ensures the Lorentz invariance of the product $K\mathcal{O}$, with $\vec{v}_i$ the velocity of particles in beam $i$. The approximation $K \approx 2c$ holds in the presence of a small half-crossing angle, $\phi$, when $|\vec{v}_i| \approx c$~\cite{Ferro-Luzzi:2012zxa}, as is the case at the \lhcb interaction point. The integration runs over all space and the time relevant to a single bunch crossing. 

This paper presents a calibration of the absolute luminosity for proton-proton (\proton\proton) collisions with a centre-of-mass energy $\sqrt{s} = 13\tev$ at the \lhcb experiment during Run 2 of the \lhc. The calibration is performed using data collected during atypical running conditions, as described in Sec.~\ref{sec:conditions}. The visible cross-section of a given counter process, $\sigma_\mathrm{vis}$, is directly measured using data collected during the calibration period. Measurements of the relative luminosity of the same process, $\mu_\mathrm{vis}$, can then be used to determine the luminosity during physics data-taking using Eq.~\ref{eq:lumi-rate}. The linearity of each counter with luminosity, and its stability against detector ageing, is verified by comparing the relative luminosities measured by counters from different subdetectors in different operational conditions. By equating Eqs.~\ref{eq:lumi-rate} and~\ref{eq:lumi-beam} the visible cross-section can be expressed as
\begin{equation}
\label{eq:xsec}
    \sigma_\mathrm{vis} \approx \frac{\mu_\mathrm{vis}}{2c N_1 N_2 \mathcal{O}}.
\end{equation}
The overlap integral is determined analytically from a model of the bunch profiles fitted to reconstructed beam-gas and beam-beam vertices~\cite{Ferro-Luzzi:2012zxa}. The interaction rate is determined by counting the fraction of empty events, and the bunch populations are measured using \lhc instrumentation~\cite{DCCT,FBCT}. 

The uncertainty on the absolute luminosity calibration represents a significant, and in some cases dominant, uncertainty on several measurements reported by the \lhcb collaboration~\cite{LHCb-PAPER-2016-021,LHCb-PAPER-2018-003,LHCb-PAPER-2018-011,LHCb-PAPER-2021-010,LHCb-PAPER-2021-037,LHCb-PAPER-2023-022}. The result presented here reduces the uncertainty on the absolute luminosity calibration used in such measurements by more than a factor of two. An internal vdM calibration with a relative uncertainty of 2\% is used in Refs.~\cite{LHCb-PAPER-2016-021,LHCb-PAPER-2018-003,LHCb-PAPER-2018-011,LHCb-PAPER-2021-010,LHCb-PAPER-2021-037,LHCb-PAPER-2023-022}. The uncertainty on the vdM calibration includes a subleading systematic uncertainty related to the relative luminosity measurements during physics data-taking which is not included in this measurement.  Additionally, the uncertainty on the BGI calibration is reduced by around a third compared to that of Ref.~\cite{LHCb-PAPER-2014-047}. The improvement is achieved primarily through a refined method to determine the beam-gas vertex resolution and estimate the associated systematic uncertainty.

\section{The \lhcb detector and data-taking conditions}
\label{sec:conditions}
The \lhcb detector is a single-arm forward spectrometer covering the pseudorapidity range $2 < \eta < 5$, designed for precision studies of the properties of particles containing \bquark\ or \cquark\ quarks. The detector consists of a tracking system, ring-imaging Cherenkov detectors for charged hadron identification, a calorimeter system, and a muon identification system. The tracking system consists of a silicon-strip detector (\velo) surrounding the interaction region, a set of silicon-strip tracking stations upstream of a dipole magnet, a further downstream set of silicon-strip and straw-tube tracking stations~\cite{LHCb-DP-2008-001,LHCb-DP-2014-002}. The \velo, used to reconstruct PVs and displaced decay vertices, is of particular relevance to the measurements presented here \cite{LHCb-DP-2014-001}. It consists of two approximately 1\m-long retractable halves, each composed of 21 modules. The nominal \proton\proton interaction region (IR) is offset in the longitudinal direction from the centre of the \velo to provide a larger acceptance for tracks travelling into the spectrometer. Each module consists of two approximately semicircular sensors which provide polar-coordinate measurements. In each pair, one sensor contains radial strips, while the other contains azimuthal strips. A pile-up system (PU) consisting of two additional modules containing only radial sensors, located upstream of the IR, is used to estimate the number of PVs in each event at the hardware trigger level. Each \velo half is retracted during injection of the \lhc beams, and then closed once the beam conditions are stable. The \velo sensors are held in a secondary vacuum which is separated from the primary beam vacuum by a radio-frequency (RF) shielding foil. The SMOG system allows a small amount of inert gas to be injected into the primary vacuum around the IR to increase the vacuum pressure from around $10^{-9}$~mbar to around $10^{-7}$~mbar and thus increase the rate of collisions between the \lhc beams and gas molecules~\cite{Barschel:2014iua}.

The beam travelling towards the \lhcb spectrometer at the IR is denoted beam~1; the other beam is denoted beam~2. The forward and backward directions are defined with respect to the direction of beam~1, \ie~forward points into the \lhcb spectrometer from the interaction point. The term `beam-gas interaction' refers to an inelastic collision between a proton in one of the \lhc beams and a gas molecule in the beam-pipe that produces a PV. The RF cavities, which accelerate the bunches in the \lhc, operate at a frequency of approximately 401\mhz. Combined with the circumference of the \lhc ring, this gives 35,640 2.5\ns-long windows defined as `RF buckets'. The buckets are further grouped into 25\ns-long periods defined as `bunch slots'~\cite{Bruning:782076}. A single RF bucket in each bunch slot is nominally filled. In each bunch crossing at \lhcb, there are four possible configurations depending on whether the bunch slot in each beam is nominally filled or empty. Colliding crossings, where both bunch slots are filled, are denoted $bb$. Crossings where only the bunch slot in beam 1 (2) is filled are denoted $be$ ($eb$). Crossings where both bunch slots are empty are denoted $ee$. Collectively, $be$, $eb$, and $ee$ crossings are referred to as noncolliding. A right-handed coordinate
system which has the $z$-axis along the beam, increasing in the forward direction, and the $y$-axis upwards along the vertical is adopted~\cite{LHCb-DP-2008-001}. The origin is defined at the nominal interaction point. This differs slightly from the conventional \lhc coordinate system; the effect of this choice is studied in Sec.~\ref{sec:results}.

The conditions of the calibration fills studied in this measurement are summarised in Table~\ref{tab:fills}. Data collected during seven calibration fills across the period 2016--18 are analysed, with either helium or neon injected around the IR using the SMOG system. In all cases, the \lhc was operating with proton beams colliding with $\sqrt{s} = 13$\tev and a nominal value of $\betastar = 24$\m at the \lhcb IR, corresponding to a typical transverse bunch width of around 100\mum. In typical physics data-taking conditions, a value of $\beta^* = 3$\m is used. The increased beam widths in calibration fills reduce the relative contribution of the \velo vertex resolution, and therefore also reduce also the associated systematic uncertainty. The \lhcb magnet was in the configuration in which the magnetic field is oriented in the negative $y$-direction for all calibration fills.

\begin{table}[tp]
    \centering
    \caption{\small Calibration fills studied in this analysis.}
    \label{tab:fills}
    \begin{tabular}{ccccc}
         \hline\hline
         Fill & Year &  Beam 1/2 bunches & Colliding at \lhcb & SMOG gas species \\
         \hline
         4937 & 2016 & 55/51 & 16 & He\\
         4945 & 2016 & 56/52 & 8 & He\\
         4954 & 2016 & 56/52 & 8 & He\\
         6012 & 2017 & 57/56 & 24 & Ne\\
         6016 & 2017 & 52/52 & 8 & Ne\\
         6864 & 2018 & 70/70 & 22 & He\\
         6868 & 2018 & 140/140 & 23 & He\\
         \hline\hline
    \end{tabular}
\end{table}

An online event selection is performed by the \lhcb trigger system, which consists of a hardware stage and two software stages. To select a sample of vertices from which the bunch profiles can be determined, the following requirements are applied. At the hardware stage, events are retained if there is sufficient activity in the calorimeter system or the PU. The first software stage performs a partial event reconstruction, including track and vertex fitting. At this stage, vertices in noncolliding bunch crossings are required to be associated with at least ten \velo tracks and to have a reconstructed longitudinal position satisfying $|z| \leq 2\m$. Vertices in colliding crossings are required either to be associated with at least ten \velo tracks and have a reconstructed position in the longitudinal range $0.25\m < |z| \leq 2\m$, or to have at least 28 \velo tracks and a reconstructed position in the longitudinal range $|z| \leq 0.25\m$. From 2017 onwards, a fraction of events in colliding crossings were randomly selected and retained if a vertex associated with at least ten \velo tracks was reconstructed in the central region. The second software stage performs a full event reconstruction including particle identification. In all cases, vertices are required to have a radial position satisfying $\sqrt{x^2 + y^2} < 4\mm$ to remove interactions originating in the detector material. Since this analysis requires only a set of reconstructed vertices, no requirements are applied at the second software stage.

Offline, a tighter radial requirement defined by $\sqrt{x^2 + y^2} < 3$\mm is imposed on each vertex to further suppress material interactions. To improve the average vertex resolution in the sample studied here and reduce the associated systematic uncertainties, beam-gas vertices are required to have more than 20 tracks and lie in the longitudinal region $250 \leq |z| < 500$\mm. Vertices that do not satisfy these two final requirements but pass all other selection requirements are utilised for systematic uncertainty studies and for measurements of the `ghost' charge fraction (see Sec.~\ref{sec:populations}). Beam-gas vertices are distinguished by requiring each vertex originating from beam~1 (2) to have no backward (forward) tracks.

To determine the mean number of interactions per crossing, bunch crossings of each type are randomly retained at a fixed frequency at the first software stage. During the calibration fills analysed here, unbiased samples of $bb$, $be$, $eb$, and $ee$ bunch crossings were retained with frequencies of 5~(40)\khz, 2~(2.27)\khz, 2~(2.27)\khz and 0.5~(0.91)\khz during 2016--17~(2018), respectively. The  `Vertex' counter, which counts the number of PVs reconstructed with coordinates satisfying ${\sqrt{x^2 + y^2} < 4\mm}$ and $-300 \leq z \leq 300$\mm, is particularly relevant to this measurement.

\section{Bunch population determination}
\label{sec:populations}
Measurements of the bunch populations at the \lhc are primarily provided by two sets of instruments: the Direct-Current Current Transformers (DCCT)~\cite{DCCT} and the Fast Beam-Current Transformers (FBCT)~\cite{FBCT}. The DCCTs provide measurements of the total current in each beam with a precision of $\mathcal{O}(0.2\%)$~\cite{Barschel:1425904}. The FBCTs have a short integration window to resolve the fraction of charge in each 25\ns bunch slot but provide less accurate measurements than the DCCTs. The population of each bunch is determined as the relative population provided by the FBCTs scaled by the total beam current provided by the DCCTs. For bunch $j$ in beam $i$ this is given by
\begin{equation}
\label{eq:populations}
    N_{i,j} = (1 - f_{\mathrm{ghost}, i}) (1 - f_{\mathrm{sat}, i}) N^\mathrm{DCCT}_i \frac{N^\mathrm{FBCT}_{i,j}}{\sum_k N^\mathrm{FBCT}_{i,k}},
\end{equation}
where $N^X_{i,j}$ represents the population of bunch $j$ in beam $i$ as measured by instrument $X$, and $f_{\mathrm{ghost}, i}$ and $f_{\mathrm{sat}, i}$ represent the `ghost' and `satellite' charge fractions in each beam, respectively, and will be discussed later. The index $k$ runs over all nominally filled bunch-crossing IDs (BCID) in the \lhc for a given fill. The ATLAS Beam Pick-up Timing System (BPTX)~\cite{BPTX} also provides sensitivity to the relative bunch populations and is utilised for systematic uncertainty studies. Both the DCCTs and FBCTs have identical `A' and `B' instruments for each beam. For both devices, the A and B systems are found to agree to the level of $\mathcal{O} (10^{-4})$ in the time periods relevant to this analysis, and this difference is thus considered negligible.

For the 2015 and 2016 data-taking years, the FBCT systems used different amplifiers for signals from odd and even BCIDs, which can potentially introduce some disagreement between the cross-section measurements for different colliding bunch pairs~\cite{FBCT_upgrade}. The BPTX system treats all BCIDs equally, and thus can be used to correct for any differences introduced by the separate electronics. Firstly, the FBCT populations are parameterised in terms of the BPTX measurements as (dropping the beam and BCID subscripts $i,j$ for brevity)
\begin{equation}
    N_\mathrm{param}^\mathrm{FBCT} = a N^\mathrm{BPTX} + b.
\end{equation}
The parameters $a$ and $b$ are determined separately per fill, beam and for even and odd BCIDs using least-squares fits to the population measurements averaged over each fill for each BCID. For each of the odd BCIDs, the mismatch is corrected by adding the difference between the even and odd parameterisations in terms of the BPTX measurements as
\begin{equation}
    N^\mathrm{FBCT}_\mathrm{odd, corr} = N^\mathrm{FBCT}_\mathrm{odd} + (N_\mathrm{even,\ param}^\mathrm{FBCT} - N_\mathrm{odd,\ param}^\mathrm{FBCT}).
\end{equation}
The FBCT measurements for even BCIDs are taken without any correction. Correcting instead the even BCIDs gives a relative difference in the measured bunch populations of around $2 \times 10^{-5}$, which is negligible for this measurement. 

Some fraction of charge within the \lhc ring can lie outside the nominally filled RF buckets. Such charges in another bucket within a nominally filled bunch slot are known as satellite charges, while those outside a nominally filled bunch slot are known as ghost charges. The DCCTs are sensitive to both contributions, while only satellite charges will contaminate measurements from the FBCTs and BPTXs.

While in principle the ghost charge contributions could be determined using the FBCT or BPTX measurements in nominally empty bunch slots, in practice neither device is sensitive enough for such a measurement. Instead, the ghost charge fractions are measured by comparing the number of beam-gas vertices produced in empty bunch crossings to that in filled bunch crossings. It is assumed that the number of reconstructed and selected beam-gas interactions per bunch crossing is proportional to the charge of the incident bunch. The ghost charge fraction can be expressed as
\begin{equation}
     f^{1(2)}_\mathrm{ghost} \approx \frac{n^{1(2)}_{ee}+n^{1(2)}_{eb(be)}}{n^{1(2)}_{be(eb)}} \frac{N^{1(2)}_{be(eb)}}{N^{1(2)}_{bb} + N^{1(2)}_{be(eb)}} \frac{1}{\varepsilon^{1(2)}_\mathrm{trigger}},
\end{equation}
where $n$ represents the numbers of reconstructed beam-gas vertices, and $N$ represents the measured bunch populations. The beam and crossing type(s) are indicated in the superscript and subscript of each parameter, respectively. Beam-gas vertices in colliding bunch crossings are excluded from this measurement to avoid possible contamination from collisions between satellite charges and protons from the filled bunches. The second ratio accounts for this exclusion by scaling the denominator of the first ratio.

\begin{figure}
    \centering
    \begin{subfigure}{0.49\linewidth}
        \centering
        \includegraphics[width=0.93\linewidth]{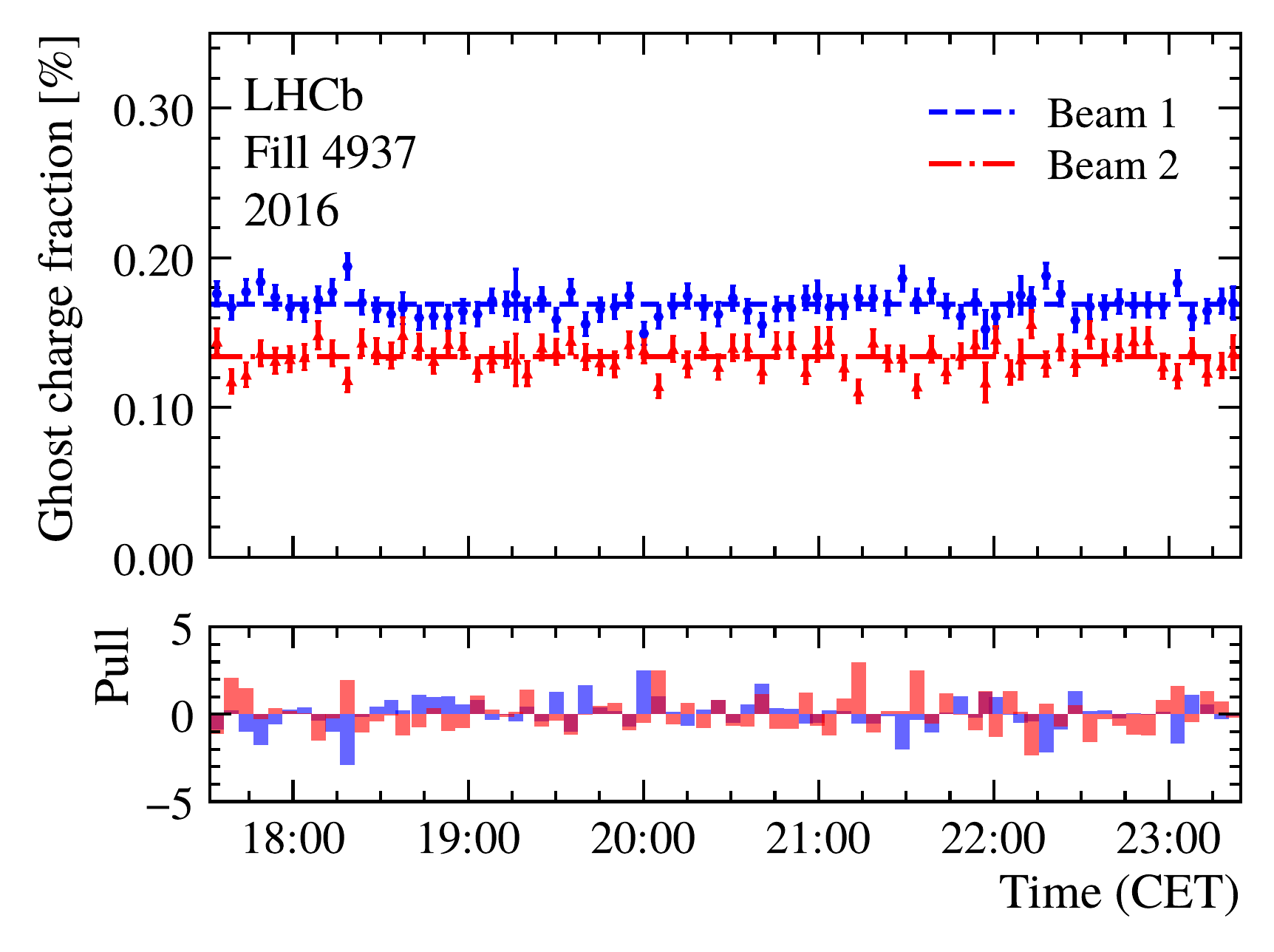}
    \end{subfigure}
    \begin{subfigure}{0.49\linewidth}
        \centering
        \includegraphics[width=0.93\linewidth]{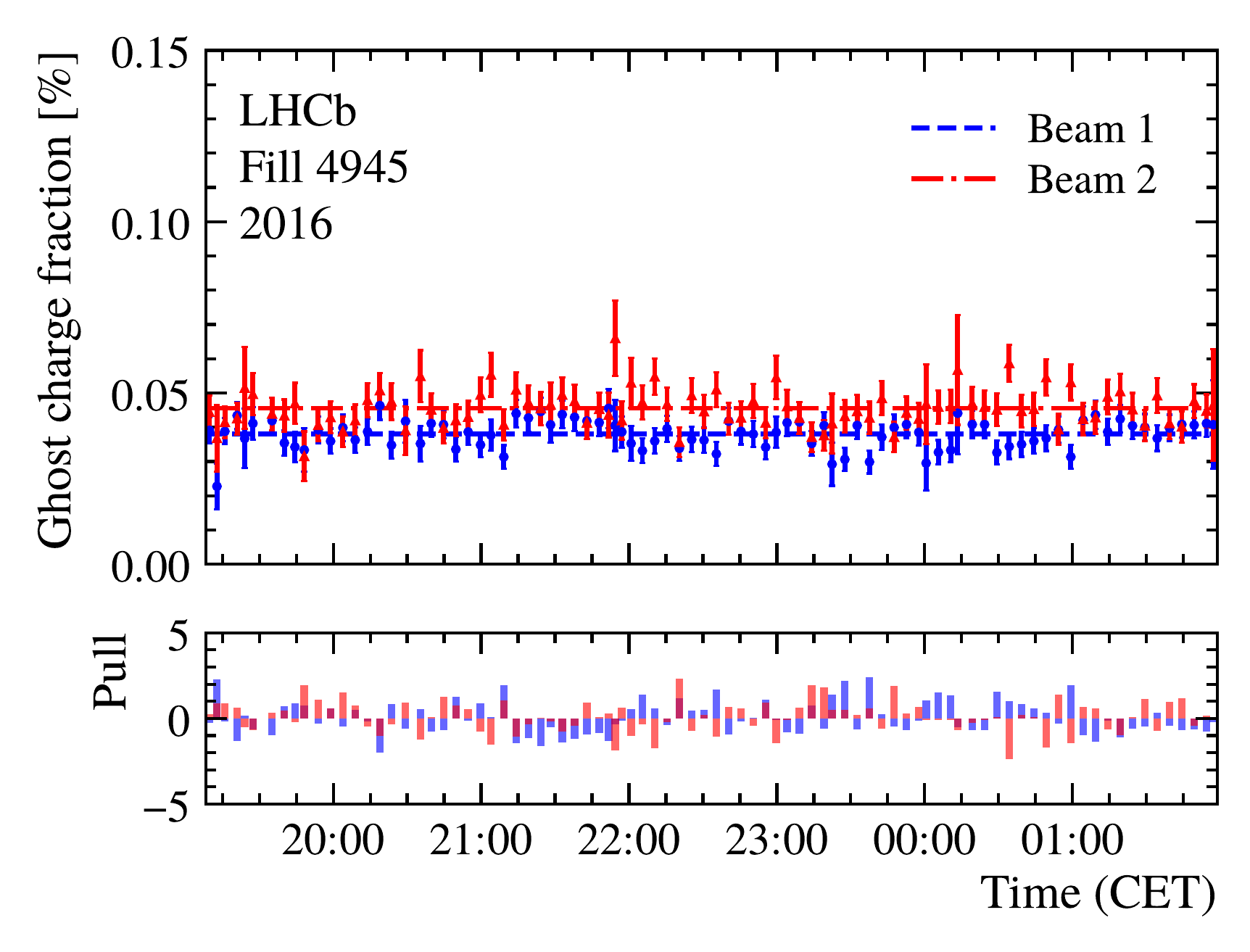}
    \end{subfigure}
    
    \begin{subfigure}{0.51\linewidth}
        \centering
        \includegraphics[width=0.93\linewidth]{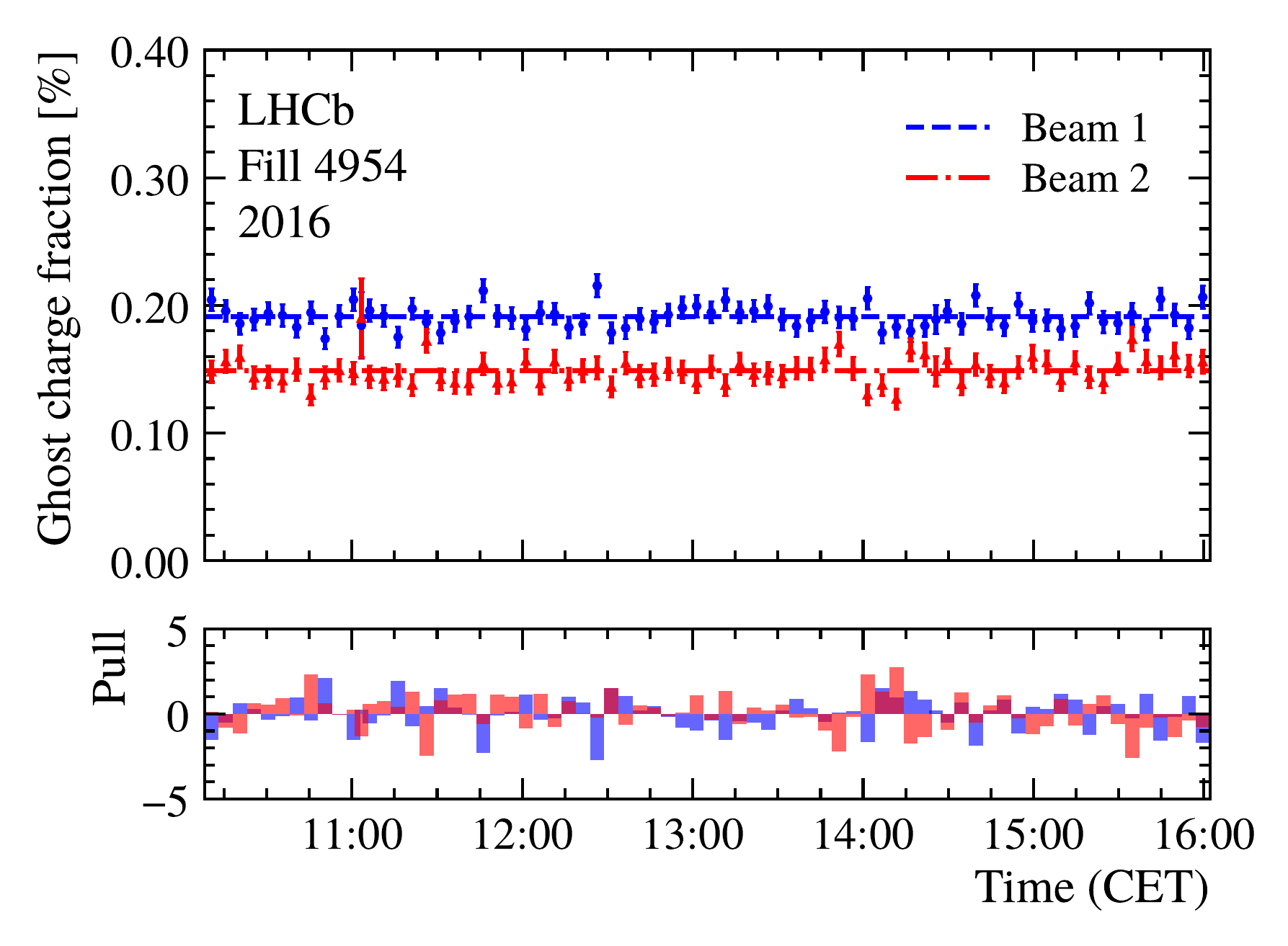}
    \end{subfigure}
    
    \begin{subfigure}{0.49\linewidth}
        \centering
        \includegraphics[width=0.93\linewidth]{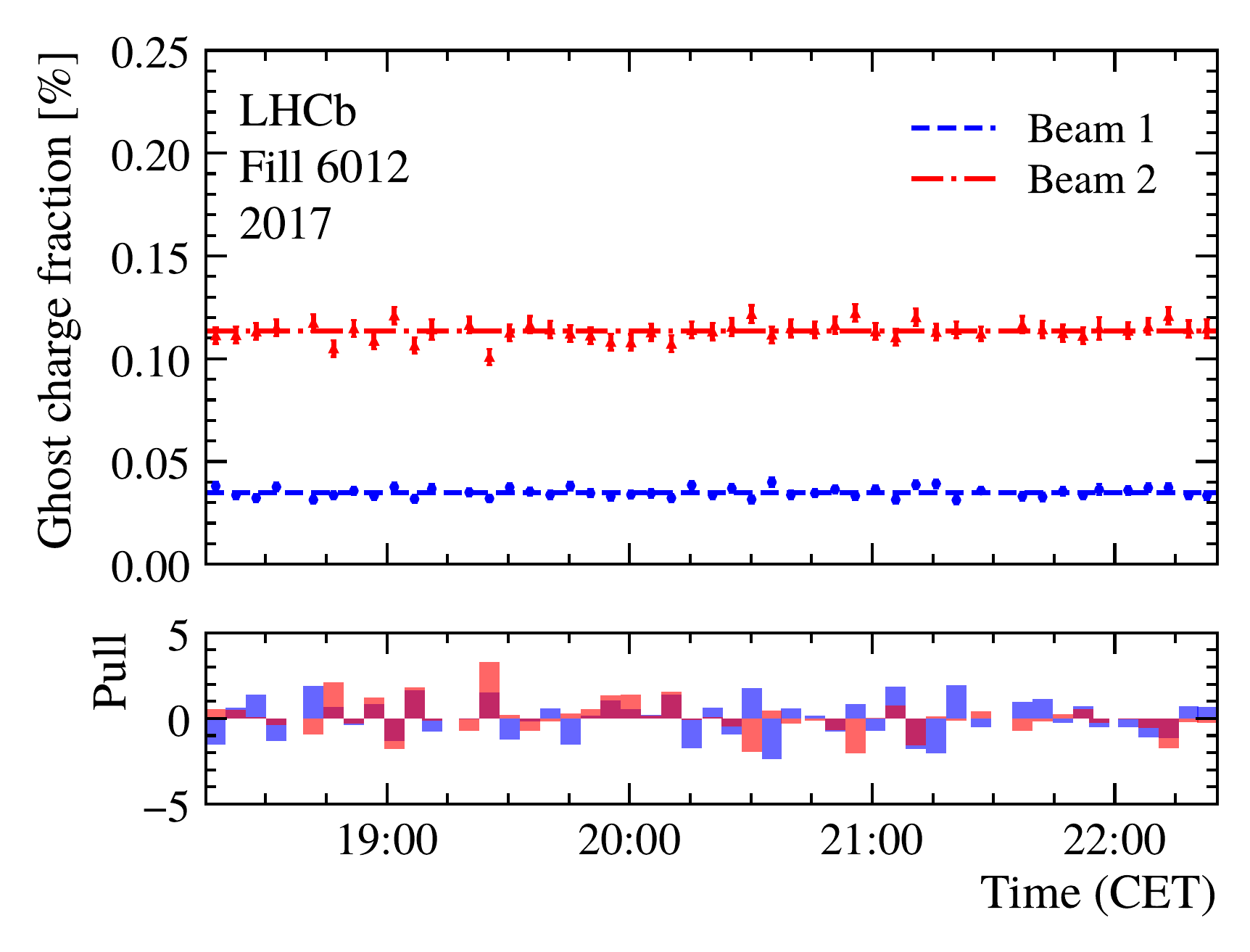}
    \end{subfigure}
    \begin{subfigure}{0.49\linewidth}
        \centering
        \includegraphics[width=0.93\linewidth]{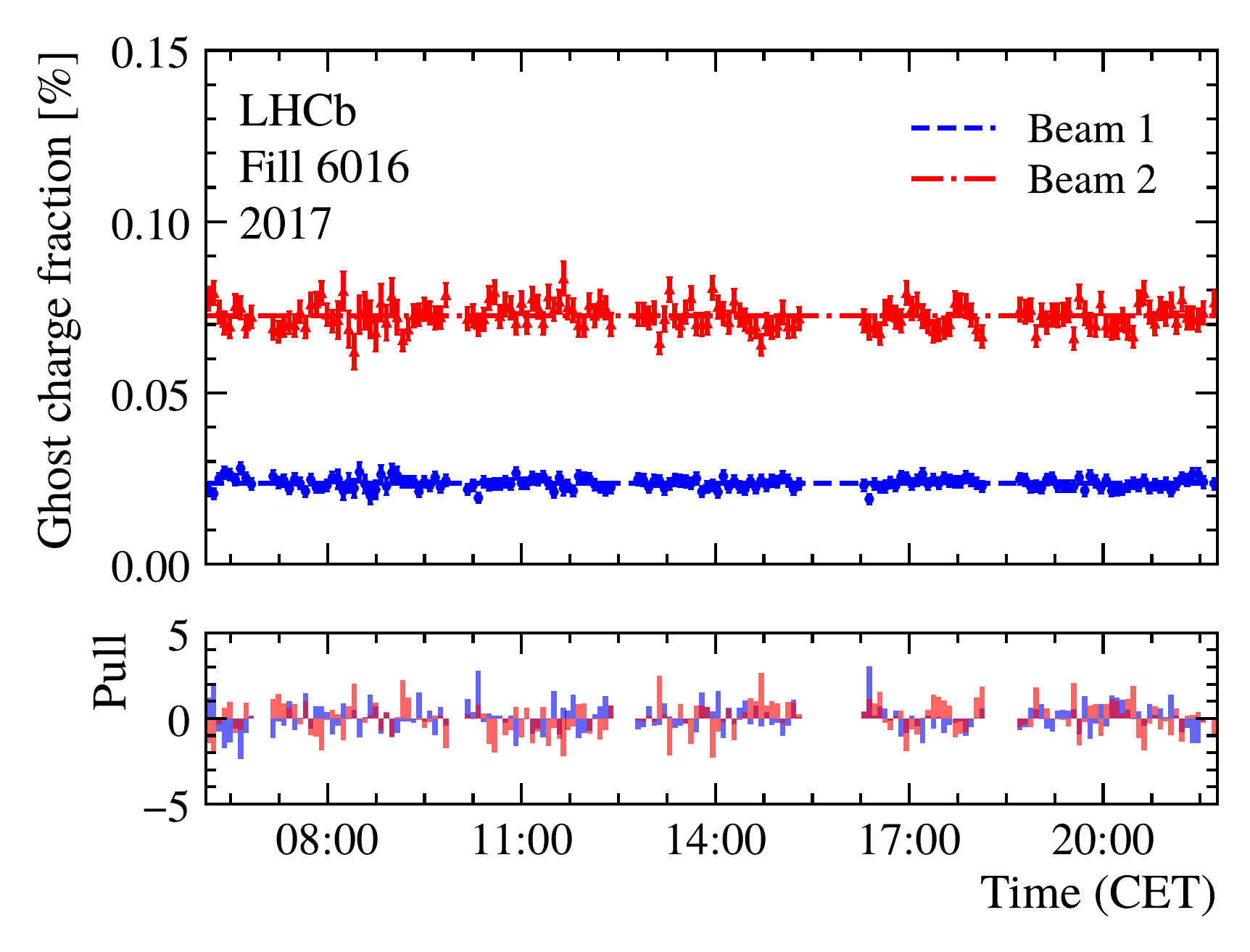}
    \end{subfigure}
    \begin{subfigure}{0.49\linewidth}
        \centering
        \includegraphics[width=0.93\linewidth]{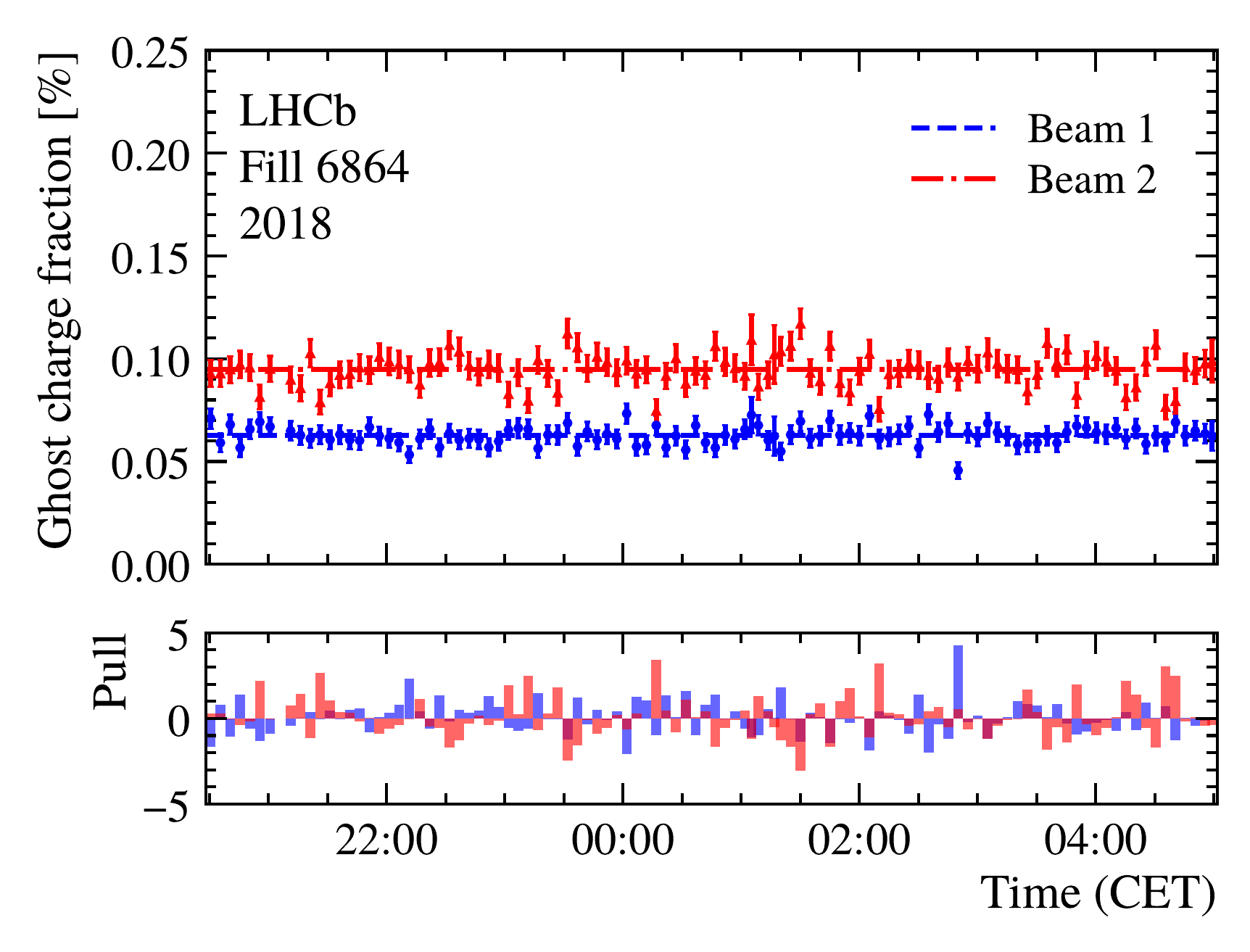}
    \end{subfigure}
    \begin{subfigure}{0.495\linewidth}
        \centering
        \includegraphics[width=0.93\linewidth]{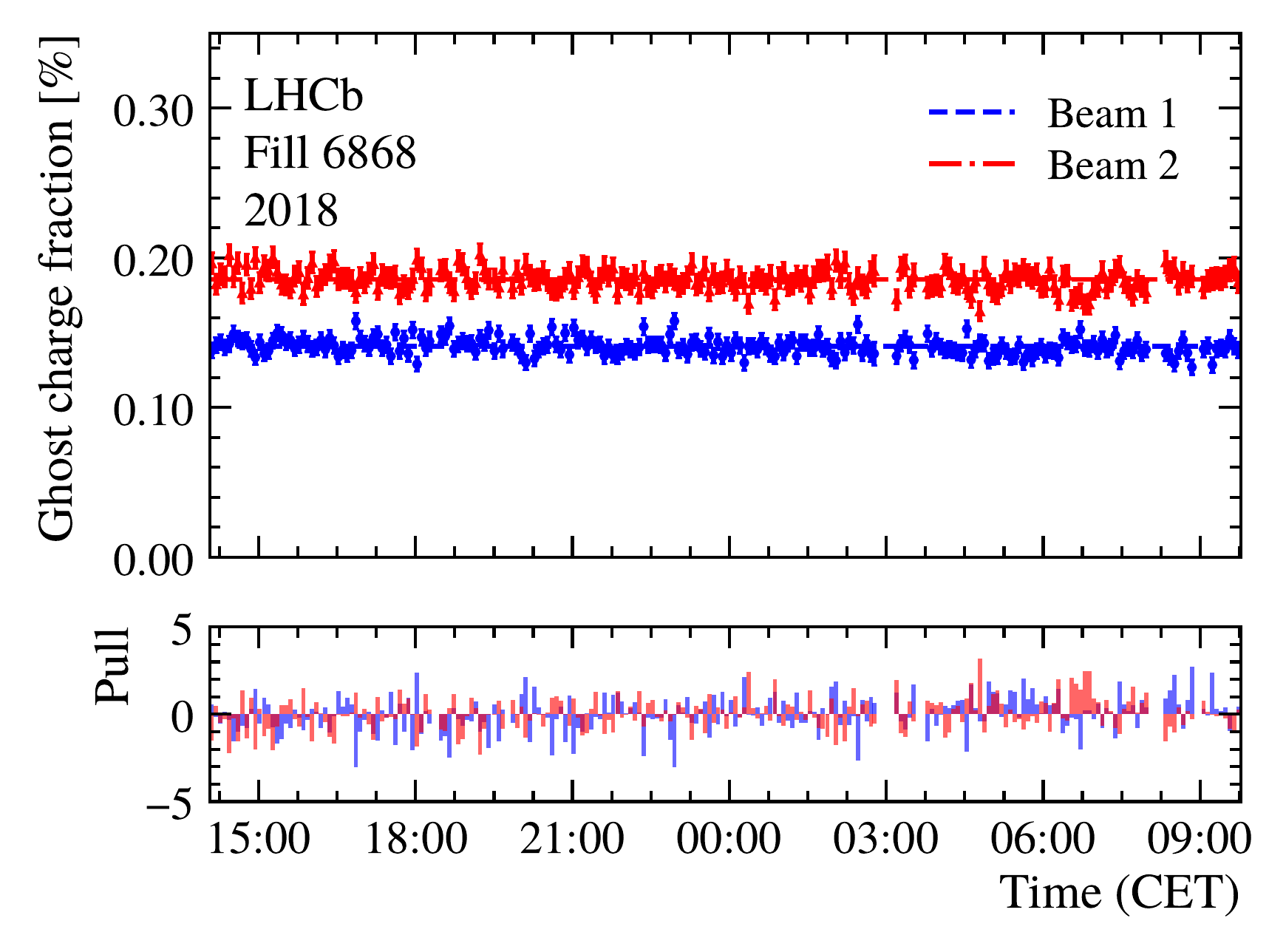}
    \end{subfigure}
    \caption{\small Ghost charge measurements for each calibration fill. Empty regions on each plot correspond to time periods that are not suitable for BGI, for example because the gas injection system was disabled, or they failed to satisfy data-quality requirements. Only statistical uncertainties are shown. For illustration purposes, the measurements are averaged over 300\sec intervals.}
    \label{fig:ghosts}
\end{figure}

While the particles in filled bunch slots are nominally contained within a single RF bucket, ghost charges can in general be distributed arbitrarily between the RF buckets. The central clock distributed to \lhcb is optimised such that the trigger efficiency is maximal on the populated RF bucket. A correction factor, $\varepsilon_\mathrm{trigger}^i$, has been measured separately for each beam to account for this effect by shifting the \lhc clock with respect to the phase of the beams and measuring the average efficiency across the bunch slot~\cite{LHCb-PAPER-2014-047, Barschel:2014iua}. This implicitly assumes that the ghost charges are uniformly distributed across the RF buckets. Additionally, interactions arising from protons in the outer RF buckets can potentially spill over between consecutive bunch slots, and potentially be double counted. To account for this, the measurement is performed both with and without removing interactions in consecutive bunch crossings. The difference between the results of the two methods is taken as a systematic uncertainty on the ghost charge fraction and their average is used as the central value. Separate trigger efficiency corrections have been measured for each approach. The ghost charge fraction is measured for each beam in 60\sec integration windows for the relevant time periods during each of the calibration fills. Figure~\ref{fig:ghosts} shows the measured ghost charge fractions for each of the calibration fills considered. The measured fraction is typically at the level of $\mathcal{O}(10^{-3})$ or smaller, and stable across the relevant measurement periods.

Measurements of the satellite charge fraction are provided by the Longitudinal Density Monitor (LDM), a single-photon counting system for synchrotron radiation produced by the \lhc beams~\cite{BSRL, BSRL2}. It is assumed that the satellite charge fraction is the same across all filled bunches, and therefore only subtracted from the DCCTs' measurements. Table~\ref{tab:satellites} shows the average satellite fraction measured for each beam in each calibration fill. The total satellite fraction is at the level of $\mathcal{O}(10^{-4})$.

\begin{table}[tpb]
    \centering
    \caption{\small Average satellite fraction measured for each beam throughout the relevant time periods of each calibration fill.}
    \label{tab:satellites}
    \begin{tabular}{ccc}
        \hline\hline
         Fill & $f_{\mathrm{sat},1}$ [$10^{-4}$] & $f_{\mathrm{sat},2}$ [$10^{-4}$]\\
         \hline
         4937 & 1.1 & 0.7 \\
         4945 & 3.0 & 2.2 \\
         4954 & 4.4 & 2.9 \\
         6012 & 0.2 & 0.8 \\
         6016 & 2.0 & 1.6 \\
         6864 & 1.7 & 2.6\\
         6868 & 5.1 & 5.8\\
         \hline\hline
    \end{tabular}
\end{table}

\section{Relative luminosity measurements}
The mean interaction rate is determined from the fraction of empty events as
\begin{equation}
\label{eq:logzero}
    \mu = -\log\left( \frac{n_\mathrm{empty}}{\ntot} \right),
\end{equation}
where $n_\mathrm{empty}$ is the number of events, from a total sample \ntot, which are defined as empty based on the lack of observation of a given counter process. The number of occurrences of the relevant process per bunch crossing is assumed to be Poisson distributed with mean~$\mu$. Higher-order corrections to ensure an unbiased estimate of $\mu$ in the case of small \ntot are found to be negligible. Biases caused by variations in $\mu$ over the relevant integration window due to the nonlinearity of Eq.~\ref{eq:logzero} are also found to be negligible. Values for $n_\mathrm{empty}$ and \ntot are 
determined from the randomly triggered events described in Sec.~\ref{sec:conditions}. The method is robust against loss of events either during data collection or offline, under the assumption that empty and nonempty events are affected in the same way. The background contribution from beam-gas interactions for each beam is estimated from $be$ and $eb$ crossings and is subtracted from the raw measured interaction rate, $\mu_\mathrm{vis}^\mathrm{raw}$. The mean interaction rate, $\mu_\mathrm{vis}$, is therefore given by
\begin{equation}
    \mu_\mathrm{vis} = \mu^\mathrm{raw}_\mathrm{vis} - N_1\langle \nu_{c,1} \rangle - N_2\langle \nu_{c,2} \rangle,
\end{equation}
where $N_i$ is the population of the relevant bunch in beam $i$, and $\langle \nu_{c,i} \rangle$ is the average specific interaction rate for filled, noncolliding crossings in beam $i$. The specific interaction rate is defined as $\nu_{c,1(2)} \equiv \mu_\mathrm{vis}^{be(eb)}/N_{1(2)}^{be(eb)}$. All quantities here, with the exception of the average specific interaction rates, refer to a single BCID, the index of which is omitted for brevity. Any baseline noise affecting the counter will be subtracted twice using this procedure, which can be corrected by adding the interaction rate measured in empty bunch crossings. For the counters studied here, this correction is found to be negligible.

\begin{figure}[tp]
    \centering
    \includegraphics[width=0.7\linewidth]{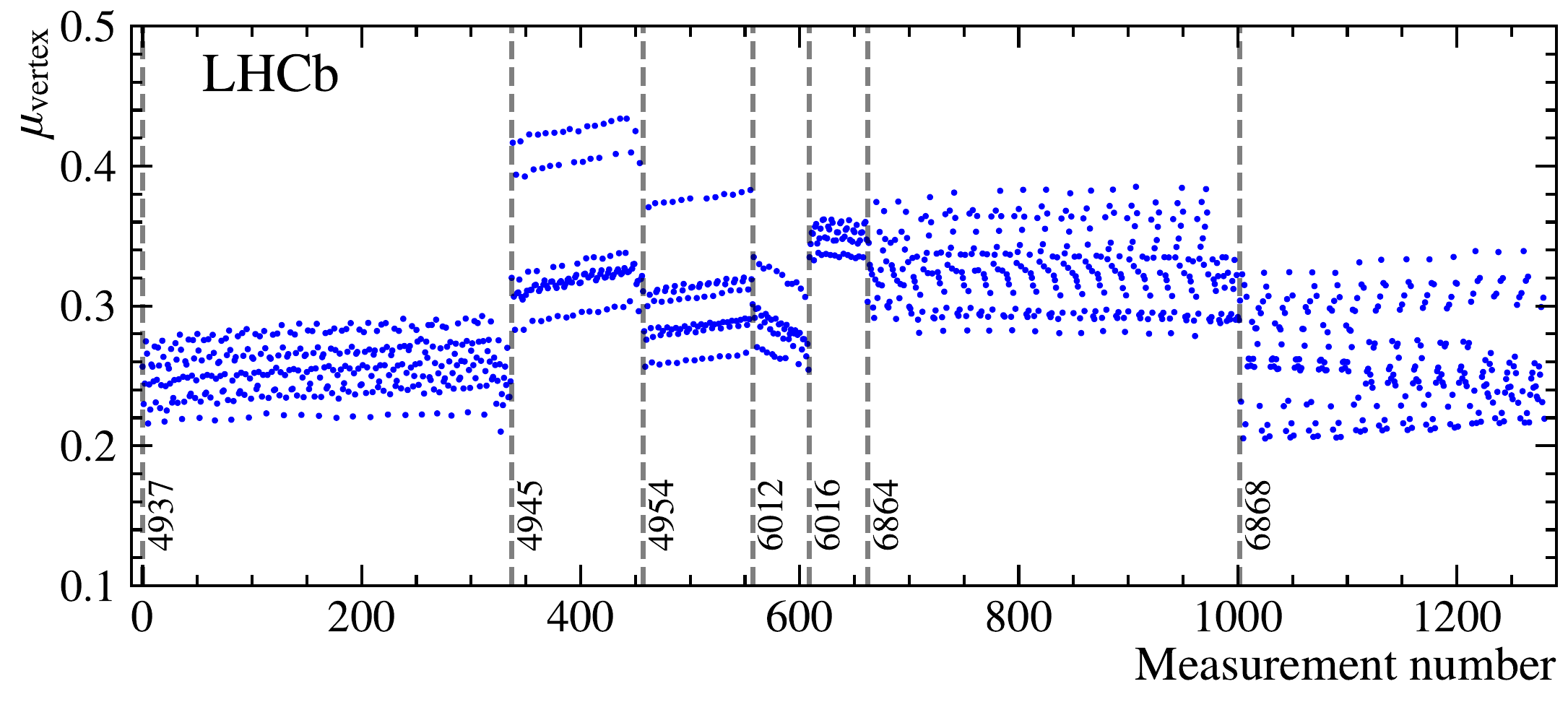}
    \includegraphics[width=0.715\linewidth]{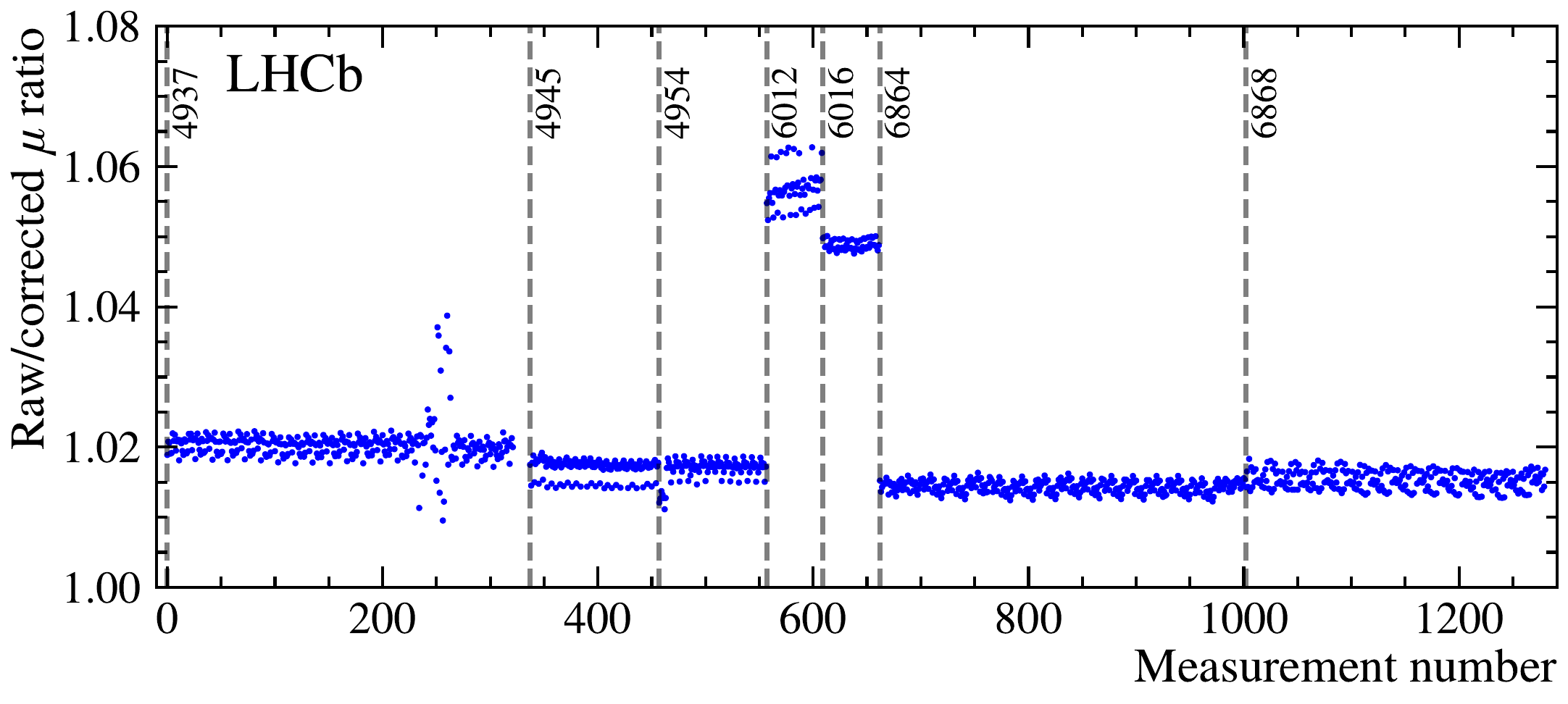}
    \caption{\small Interaction-rate measurements for the Vertex counter shown (top) after beam-gas background subtraction and (bottom) as the ratio before and after beam-gas background subtraction. Each data point corresponds to a single integration window and BCID. The data points are sorted chronologically, and then by BCID. The dashed vertical lines indicate the boundaries between the different fills included in this measurement. Within each fill, the appearance of several distinct trends corresponds to the different BCIDs.}
    \label{fig:mu-background-corr}
\end{figure}

Figure~\ref{fig:mu-background-corr} shows the distribution of interaction rate measurements for the Vertex counter after the background subtraction, and the ratio of the measurements before and after the correction. The background correction is significantly larger in fills during which Ne, as opposed to He (see Table~\ref{tab:fills}), was injected using SMOG. The larger fluctuations during part of fill 4937 are statistical in nature due to a set of short integration periods related to the data-taking conditions, and do not bias the overall measurement.

\begin{figure}[tp]
    \centering
    \begin{subfigure}{0.49\linewidth}
        \includegraphics[width=\linewidth]{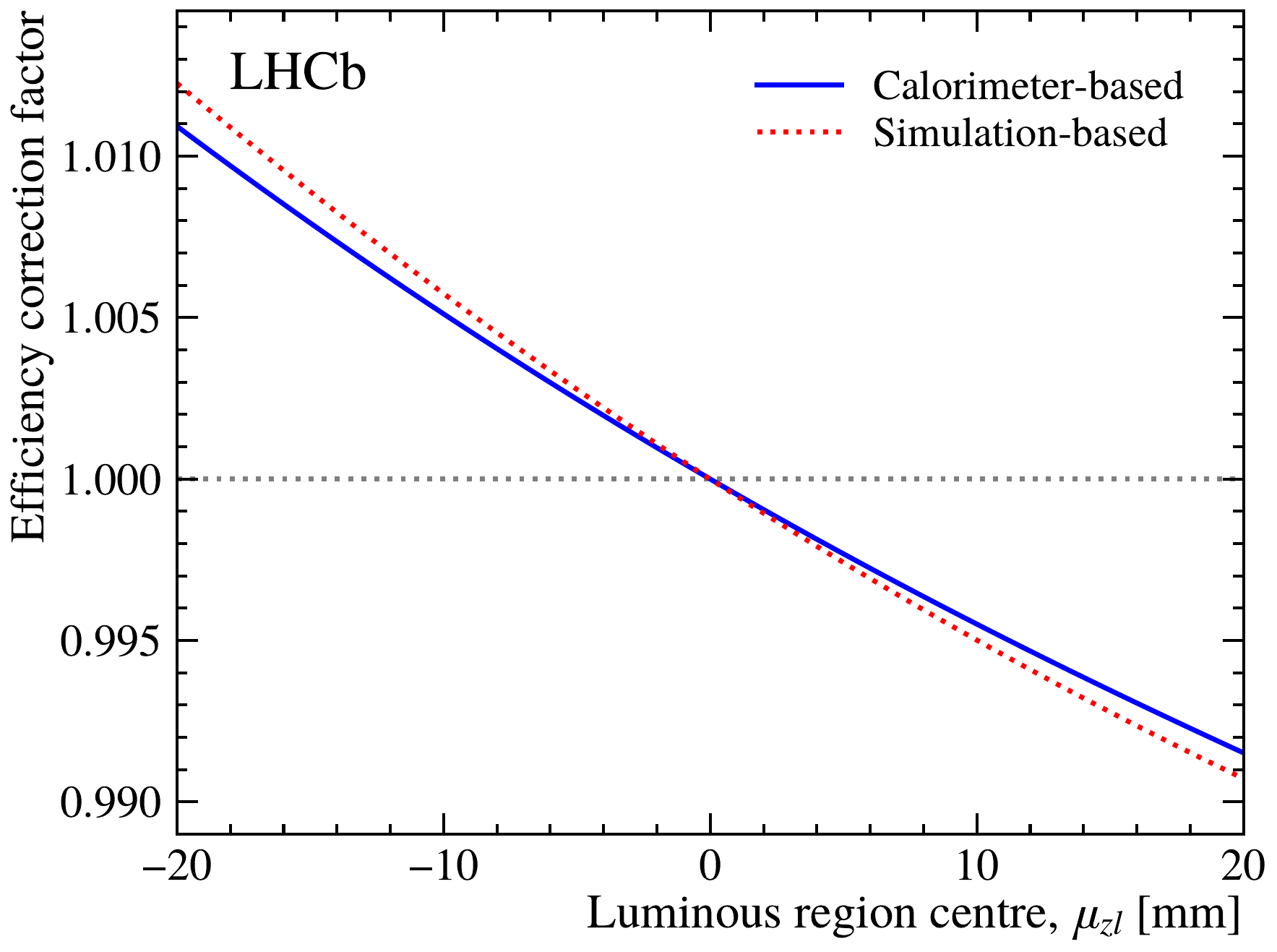}
    \end{subfigure}
    \begin{subfigure}{0.49\linewidth}
        \includegraphics[width=\linewidth]{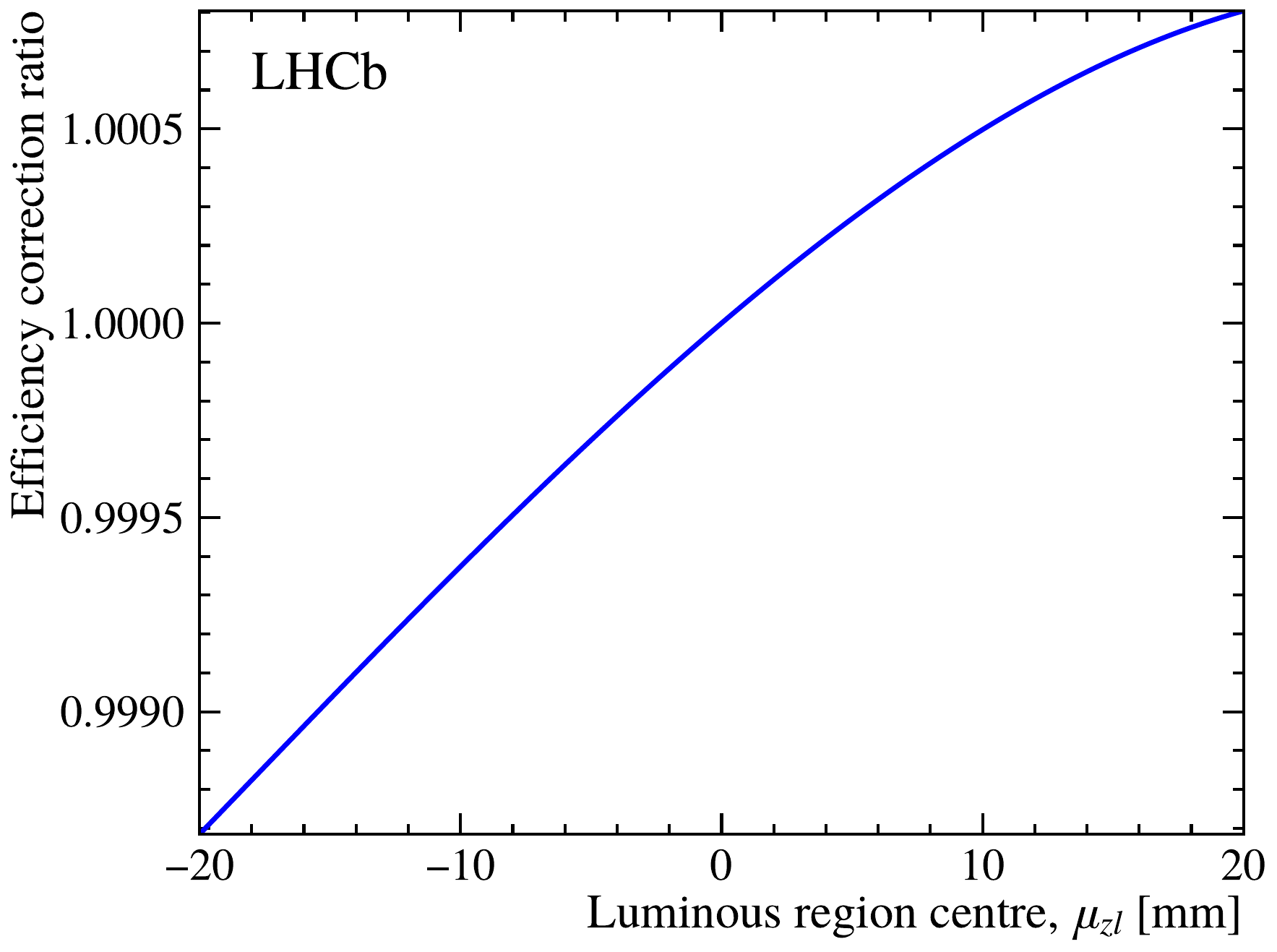}
    \end{subfigure}
    \caption{\small (Left) Correction factor applied to the Vertex counter interaction rates as a function of luminous region centre, assuming $\lambda_{zl}=50$\mm, and (right) the ratio of the corrections calculated using calorimeter counters to those calculated using simulation.}
    \label{fig:mu-z-dependence}
\end{figure}

Due to the arrangement of the modules within the \velo, the vertex reconstruction efficiency changes as a function of longitudinal position. Thus, the Vertex counter is sensitive to the position and shape of the luminous region. To correct for this, the measured values of each counter are normalised to the fixed reference point of a Gaussian-shaped luminous region centred at $z=0$, with a width of 50\mm~\cite{LHCb-PAPER-2014-047}. For the purposes of this correction, the width and centre of the luminous region are measured without correcting for their dependence on the reconstruction efficiency. Since the same procedure is used to determine these parameters during physics data-taking, the effect of this bias on the final luminosity is found to be negligible. The correction factor is calculated from the reconstruction efficiency determined from a simulated sample of minimum bias events~\cite{LHCb-PAPER-2014-047}. An alternative estimation of the relative efficiency variations of each \velo-based counter is derived from data by employing measurements from the calorimeters, which are largely insensitive to the position and shape of the luminous region. The efficiency correction as a function of the position of the luminous region, assuming a width of $\sigma_{zl} = 50$\mm, is shown in Fig.~\ref{fig:mu-z-dependence} for the Vertex counter, alongside the ratio of the corrections derived using the two methods. The difference between the two methods is typically less than $0.1\%$.

\section{Overlap integral determination}
\label{sec:overlap}

The overlap integral is determined using the results of a set of fits to the distribution of vertices produced in beam-gas and beam-beam interactions. Firstly, the half-crossing angle is determined once for each data-taking run by fitting the distribution of beam-gas vertices.\footnote{Throughout Secs.~\ref{sec:overlap} and~\ref{sec:results}, `run' refers to an internal unit of data-taking at \lhcb, typically corresponding to one hour.} Secondly, the bunch profiles are extracted independently for each colliding bunch pair from a fit to vertices collected in 15 minute integration periods. Some measurements are taken over shorter periods due to short runs or run lengths that differ from a multiple of 15 minutes; such measurements are nonetheless included in the analysis. Finally, the longitudinal bunch distributions are determined for each BCID from a fit to the sample of beam-beam interactions collected during the same integration windows. This procedure assumes that the bunch densities can be factorised into transverse and longitudinal components, $\rho_i(\vec{x}) = \rho^\perp_i(x, y) \rho^\parallel_i(z)$; it is not assumed that  the transverse distributions are factorisable. All fits, with the exception of those to determine the half-crossing angle, are implemented as unbinned maximum-likelihood fits~\cite{roofit}.

In the case where the bunch profile along each transverse axis is represented by a single Gaussian distribution, the transverse shapes are factorisable and the overlap integral is given by~\cite{Ferro-Luzzi:2012zxa}
\begin{equation}
    \label{eq:overlap}
    \mathcal{O} = \frac{
    \exp{(-\Delta x^2/2\Sigma_x^2})\, 
    \exp{(-\Delta y^2/2\Sigma_y^2)}
                    }
    {4\pi c \Sigma_x \Sigma_y},
\end{equation}
where $\Delta a$ is the transverse offset in the $a$-axis for $a = x,y$, defined as the difference between the centres of the Gaussian distributions representing the bunches from each beam, evaluated at the crossing point, $z_\mathrm{RF}$, which is defined in this section. The parameters $\Sigma_x$ and $\Sigma_y$ are given by
\begin{equation}
\begin{aligned}
    \Sigma_x^2 &\approx 2\lambda_x^2 + 2\phi_x^2 \lambda_z^2, \\
    \Sigma_y^2 &\approx 2\lambda_y^2 + 2\phi_y^2 \lambda_z^2 \approx 2\lambda_y^2,
\end{aligned}
\end{equation}
where $\lambda_a$ is defined by $2\lambda_a^2 \equiv \lambda_{a,1}^2 + \lambda_{a,2}^2$ and $\phi_a$ is the half-crossing angle of the two beams in the $az$-plane. The final approximation is valid for the conditions of all calibration fills described in this measurement, where $\phi_y \sim 0$. The half-crossing angle is defined as $\phi_a \approx (m_{2,a} - m_{1,a})/2$ for the $a$-axis, where $m_{i,a}$ is the gradient of beam $i$ in the $az$-plane. All parameters here, with the exception of the half-crossing angles and gradients, are specific to an individual BCID, whose index is suppressed for brevity. A nonfactorisable description of the transverse bunch profile can be achieved by constructing a model from the weighted sum of many such individually factorisable terms, where the overlap integral can then be calculated simply as the weighted sum of the individual overlap integrals.

\begin{figure}[tbp]
    \centering
    \includegraphics[width=0.7\linewidth]{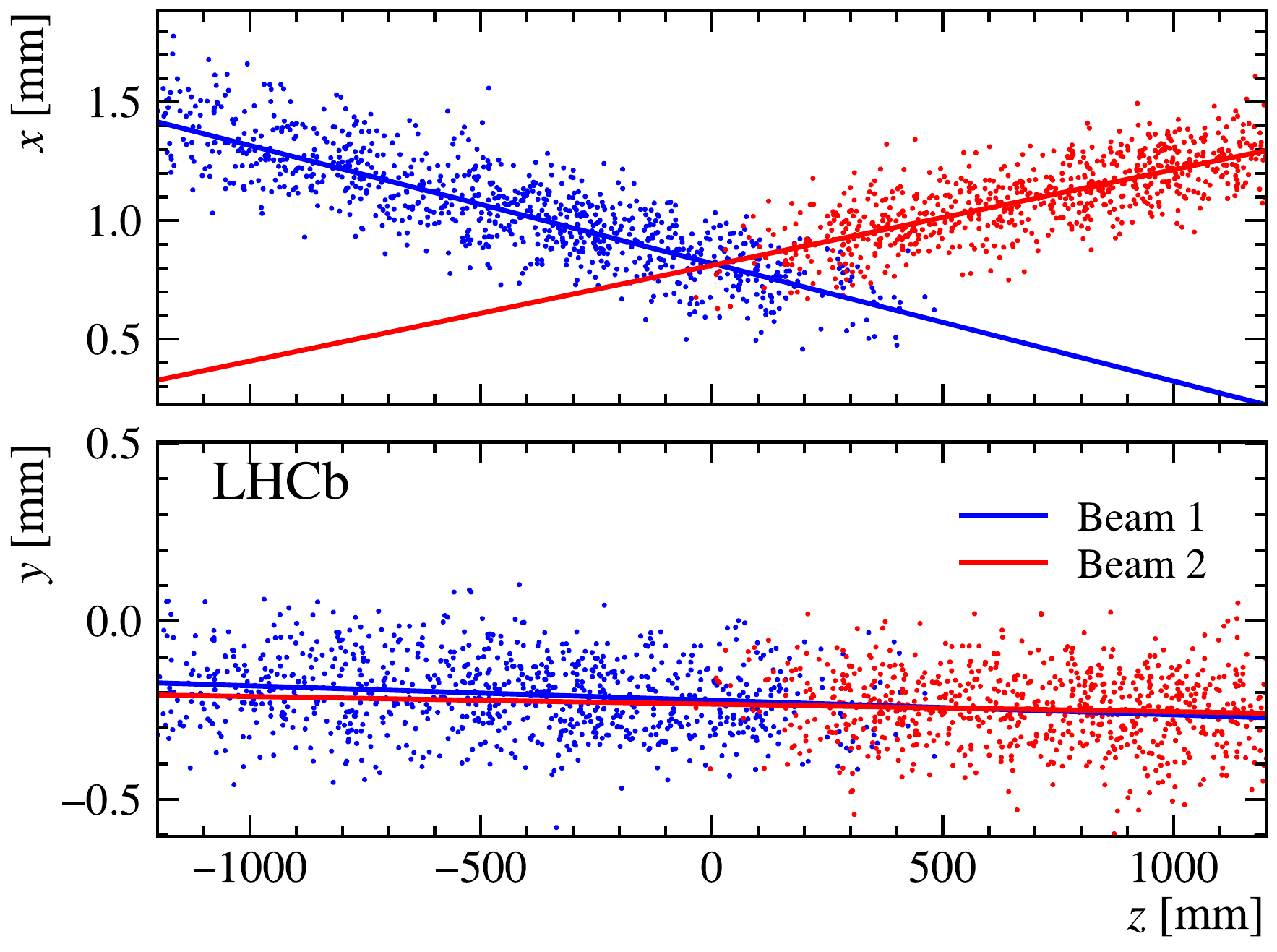}
    \caption{\small Example distribution of reconstructed beam-gas vertices in the $xz$ and $yz$ planes from Fill 4937. A set of linear fits, from which the half-crossing angles of the beams in both planes are determined, are shown.}
    \label{fig:example-crossing}
\end{figure}

The directions of the two beams in each of the transverse axes are measured from a pair of least squares fits of a linear slope to the distribution of beam-gas vertices collected during each data-taking run, as illustrated in Fig.~\ref{fig:example-crossing}. The lever arm of the fit is increased by integrating the sample of vertices across all bunches from all crossing types. The net half-crossing angle at the \lhcb IR was nominally 450~$\mu$rad in the $x$-axis and zero in the $y$-axis for all of the calibration fills used in this measurement.

The vertex resolution depends strongly on the number of tracks entering the PV fit. There is an additional dependence on the longitudinal position of the vertex, which is likely to be related to the position of the \velo modules and the opening angles between tracks entering the vertex fit. The vertex resolution is derived from data using a process known as the `split-vertex' method~\cite{LHCb-PAPER-2014-047}. The tracks from each vertex are randomly divided into two subsets and a new `split' vertex is fitted from each subset. An equal composition of tracks in each split vertex from the left and right halves of the \velo is enforced at the splitting stage, up to possible differences when splitting an odd number of tracks. It is assumed that correlations between the offsets of each split vertex from the true vertex position and the true vertex position can be neglected, as can correlations between the offsets themselves. It follows that the variance of the half-difference between the two split vertices is equal to the variance of the resolution distribution affecting their average position. The distribution of average split vertex positions will have the same bunch profile as the fully reconstructed vertex positions, albeit convolved with a different resolution distribution. 

The transverse vertex resolution is determined directly from each subsample of beam-gas vertices for each colliding bunch pair. A single-Gaussian model does not reproduce the distribution well, since each sample contains a mixture of vertices with different resolutions, depending on the parameters of the given vertex. Since the resolution depends on parameters of the vertex which are common to both axes, the product of a pair of one-dimensional sum of Gaussian components will contain unphysical cross-terms mixing the narrow and wide components which have a good resolution in one axis, but a poor resolution in the other. Thus, the components of the model in each axis are assigned an index, with each component forced to have a larger width parameter than the previous index. The two-dimensional model used in the fit is constructed as a weighted sum over products of terms from both axes sharing a common index. The fitted parameters of the resolution model are then used as an input to the transverse bunch profile fits. The vertex resolution in the longitudinal direction is neglected since the typical width of the luminous region is of the order of~50\mm, while the longitudinal resolution is around~0.1\mm.

\begin{figure}[tbp]
    \centering
    \includegraphics[width=\linewidth]{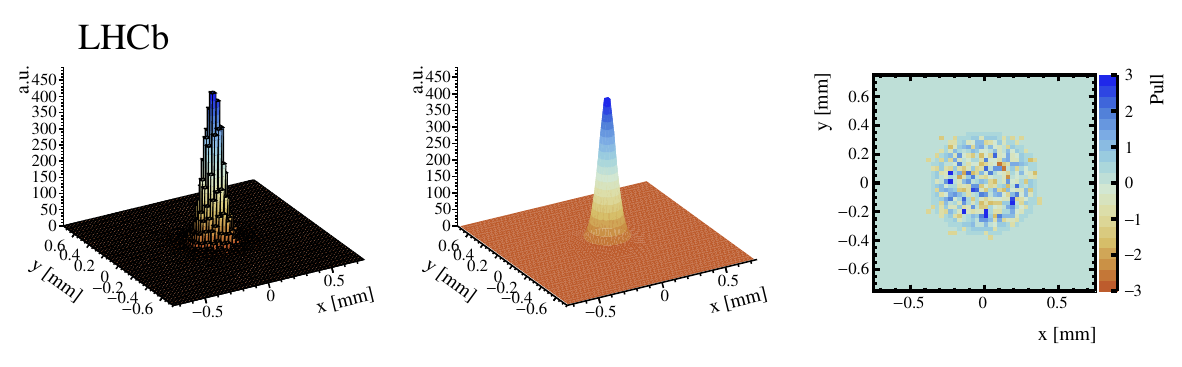}
    \includegraphics[width=\linewidth]{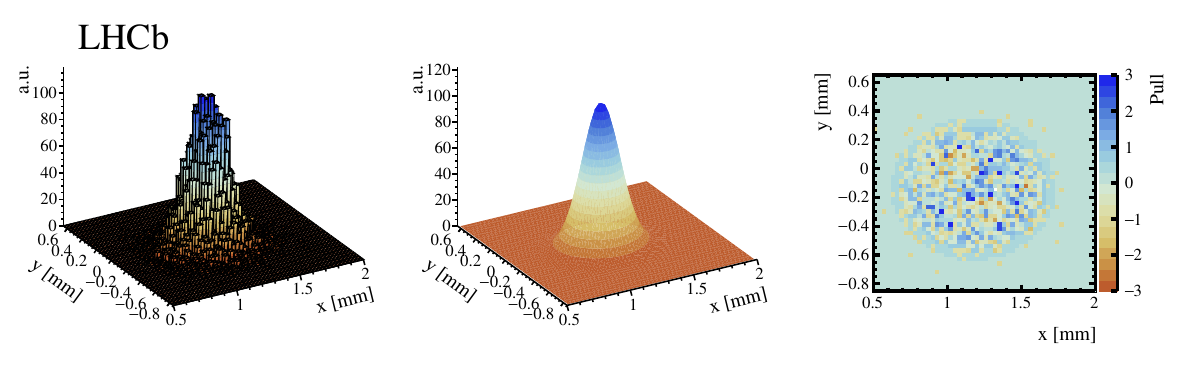}
    \caption{\small Example fit of the transverse bunch profile for beam 1, for a single BCID from fill 6012. The (left) data, (middle) fit model, and (right) pull distribution are shown. The results of the fits to determine the (top) transverse beam-gas resolution shape and (bottom) transverse bunch profile are shown. To highlight the relative size of the vertex resolution, both fits are shown with the same length of axes.}
    \label{fig:example-fit}
\end{figure}

\begin{figure}[tbp]
    \centering
    \includegraphics[width=0.7\linewidth]{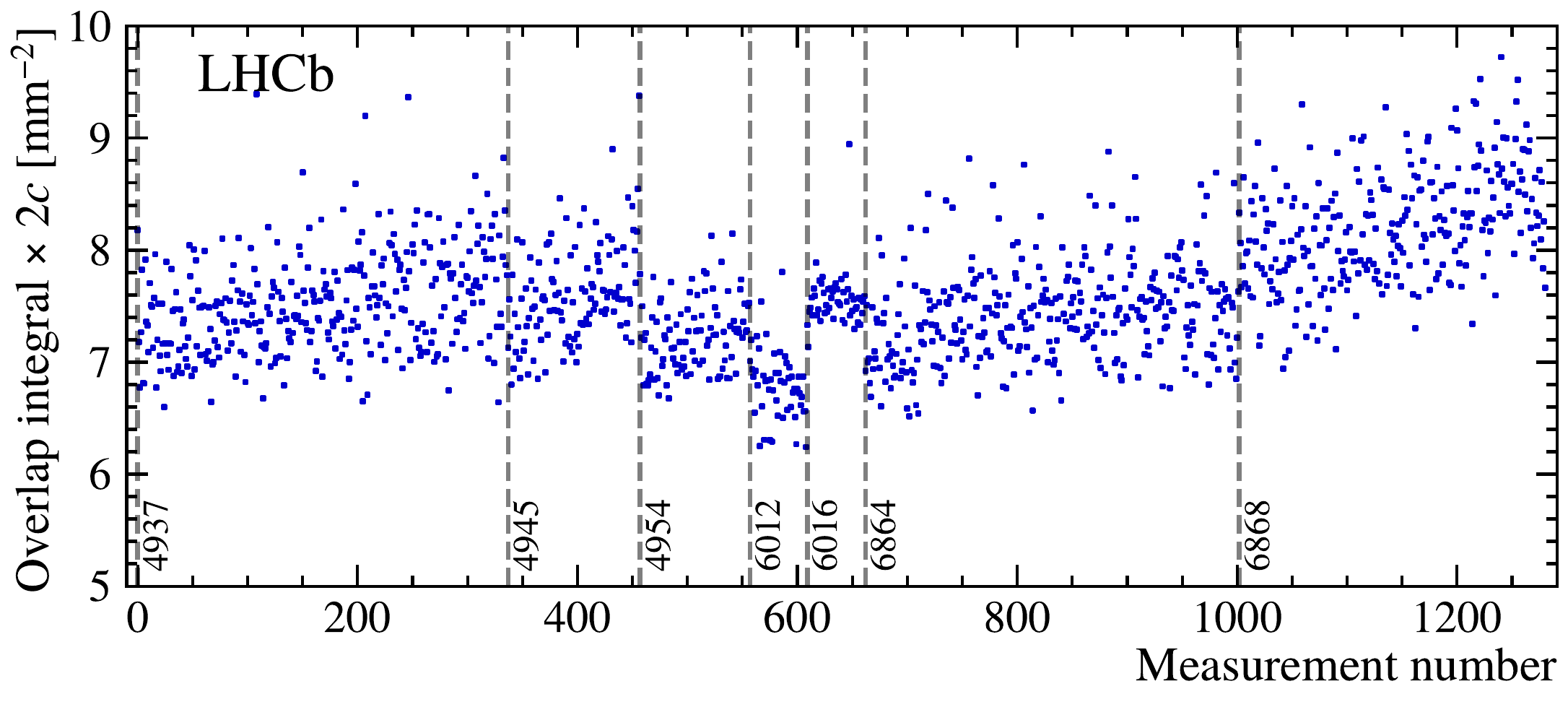}
    \caption{\small Overlap integral measurements assuming triple Gaussian bunch profiles across all calibration fills. Each data point corresponds to a single integration window and BCID. The data points are sorted chronologically, and then by BCID. The dashed vertical lines indicate the boundaries between the different fills included in this measurement. }
    \label{fig:overlap-nf}
\end{figure}

For each colliding-bunch pair the profile is determined from a fit to the sample of beam-gas vertices in each integration window. The reconstructed transverse coordinates of each vertex are projected to the centre of the longitudinal range using beam slopes fitted for the crossing angle measurements. The bunch profiles are described using a two-dimensional parameterisation that allows for nonfactorisability of the shape in the transverse axes. The bunch profiles are modelled using a sum of Gaussian products, which can be written~as
\begin{equation}
\label{eq:profile-model}
    \rho(x, y) = \sum_{n} w_n g(x\, |\, \mu_x^n, \lambda_x^n) g(y\, |\, \mu_y^n, \lambda_y^n),
\end{equation}
where $w_n$ is a weight parameter, and $\mu_a^n$ and $\lambda_a^n$ represent the mean and width of the $n$-th Gaussian function for the $a$-axis. Such a model is nonfactorisable by construction, as long as there is more than one nonzero weight and nondegenerate width, since cross terms between the different components in each axis are ignored. According to the underlying beam dynamics, the transverse profile must be symmetric about the bunch centre after projecting onto any given transverse axis~\cite{betatron-osc}. This is achieved by allowing the parameters associated with the outer Gaussian components for each axis to vary within the constraints that they share the same weight and width, and that their centres deviate from the first Gaussian function equally, but in opposite directions. A model consisting of five Gaussian products is used in the baseline analysis procedure. The model therefore has nine floating parameters: a core mean, $\mu_{x/y}$, and width, $\lambda_{x/y,0}$, for each axis; a single outer width, $\lambda_{x/y,1}$, and offset parameter, $\delta_{x/y}$, for each axis; and a single weight, $w$. Additional weights are fixed by the symmetry and normalisation requirements. More explicitly, this can be written as
\begin{align}
\label{eq:baseline-model}
    \rho(x,y) = w &g(x|\mu_x, \lambda_{x,0}) g(y|\mu_y, \lambda_{y,0}) \nonumber \\
    + \left[(1-w)/4 \right][
    &g(x|\mu_x + \delta_x, \lambda_{x,1}) g(y|\mu_y + \delta_y, \lambda_{y,1})  \nonumber\\
    + &g(x|\mu_x + \delta_x, \lambda_{x,1}) g(y|\mu_y - \delta_y, \lambda_{y,1}) \\
    + &g(x|\mu_x - \delta_x, \lambda_{x,1}) g(y|\mu_y + \delta_y, \lambda_{y,1})
     \nonumber \\
    + &g(x|\mu_x - \delta_x, \lambda_{x,1}) g(y|\mu_y - \delta_y, \lambda_{y,1})]. \nonumber
\end{align}
For each measurement, the underlying bunch profile is convolved analytically with the resolution model described above. Figure~\ref{fig:example-fit} shows an example fit to the transverse bunch profile for one beam, BCID, and integration period. Figure~\ref{fig:overlap-nf} shows the overlap integral measured using the baseline bunch profile model.

No fine timing information is available for vertices reconstructed in the \lhcb detector and therefore beam-gas vertices provide no sensitivity to the longitudinal bunch profiles. However, the longitudinal distribution of beam-beam vertices is related to the longitudinal profiles of the individual bunches and thus provides sensitivity to their convolved bunch lengths. The crossing point, $z_\mathrm{RF}$, is defined as the longitudinal position at which the longitudinal beam centres meet. This can be displaced from the centre of the luminous region, $\mu_{zl}$, in the presence of an offset of the beams in the transverse plane and a nonzero crossing angle. A temporal offset of the bunches will produce an effect that is equivalent to such a spatial offset. Assuming Gaussian bunch profiles in both the longitudinal and transverse directions, the mean position of the luminous region can be written as~\cite{Ferro-Luzzi:2012zxa}
\begin{equation}
\label{eq:lumi-mu}
    \mu_{zl} \approx z_\text{RF} - \frac{\lambda_z^2 - \lambda_x^2}{2\Sigma_x^2} \phi_x \Delta x,
\end{equation}
where it is assumed that the individual bunch widths in the $x$-direction are approximately equal. The width of the luminous region can be written as
\begin{equation}
\label{eq:lumi-width}
    \lambda_{zl} \approx \left( \frac{2}{\lambda_z^2} + \frac{2\lambda_x^2 \phi_x^2}{\lambda_{x,1}^2 \lambda_{x,2}^2} \right)^{-\frac{1}{2}}.
\end{equation}
In both cases the absence of a crossing angle in the $y$-direction is assumed. A determination of the longitudinal bunch profile parameters therefore requires the transverse profiles as an input. Assuming factorisation of the longitudinal and transverse directions, the luminous region profile, $\rho_{bb}(z)$, can be constructed from the sum of all possible combinations of the individual Gaussian components from each beam. The weight of each individual component is proportional to its contribution to the total overlap integral. This can be written as
\begin{equation}
    \label{eq:bb-profile}
    \rho_{bb}(z) = \sum_{i,j} w^1_i w^2_j \mathcal{O}_{i,j}\, g_{i,j}(z),
\end{equation}
where the indices $i$ and $j$ run over the transverse components described in Eq.~\ref{eq:profile-model} from beam 1 and 2, respectively, and $\mathcal{O}_{i,j}$ represents the overlap integral, computed using Eq.~\ref{eq:overlap}, of the $i$-th single-Gaussian component from beam 1 and the $j$-th single-Gaussian component from beam 2. The term $g_{i,j}(z)$ represents a Gaussian distribution with mean and width given by Eqs.~\ref{eq:lumi-mu} and~\ref{eq:lumi-width}, respectively, taking the transverse parameters from the relevant single Gaussian components. The parameters $z_\mathrm{RF}$ and $\lambda_z$ are shared between all individual components, and are determined from an unbinned maximum-likelihood fit of the model described in Eq.~\ref{eq:bb-profile} to the  $z$-distribution of the luminous region. Each vertex is assigned a weight in the fit to correct for the effect of a nonuniform reconstruction efficiency in the longitudinal direction, which is determined from simulation~\cite{LHCb-PAPER-2014-047}.

The total overlap integral for each BCID in each integration window is calculated analytically as the weighted sum of the individual overlap integrals, as given in Eq.~\ref{eq:overlap}, from each possible combination of the single-Gaussian components describing each beam:
\begin{equation}
    \mathcal{O} = \sum_{i,j} w_i^1 w_j^2 \mathcal{O}_{i,j},
\end{equation}
where the weight for each term is given by the product of the weights for each of the relevant components.

\section{Results and systematic uncertainties}
\label{sec:results}
The cross-section is calculated for each colliding BCID in each integration period based on the methods described in the previous sections, as given in Eq.~\ref{eq:xsec}. The mean interaction rate is evaluated in four minute intervals, which are then averaged over the relevant integration window. The overlap integral is calculated as described in Sec.~\ref{sec:overlap}. The bunch populations are calculated as given in Eq.~\ref{eq:populations}. The ghost charge measurements, nominally in intervals of 60\sec, are averaged over the relevant integration period, as are the satellite charge measurements.

\begin{figure}[tbp]
    \centering
    \includegraphics[width=0.5\linewidth]{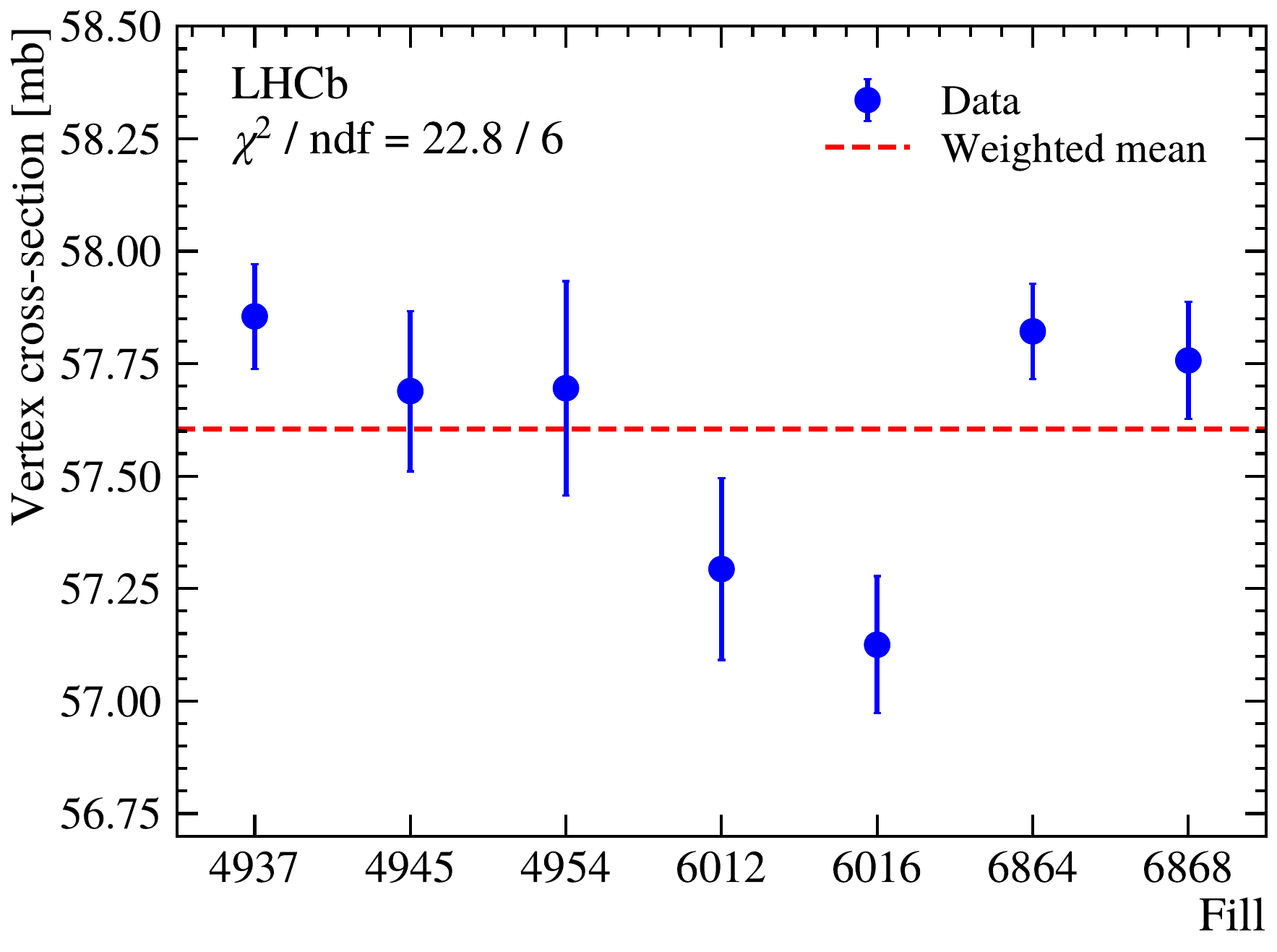}
    \caption{\small Measured cross-section of the Vertex counter per fill, with statistical uncertainties. A fit of the average cross-section across all fills is also shown.}
    \label{fig:xsec-perfill}
\end{figure}

\begin{figure}[tbp]
    \centering
    \includegraphics[width=0.7\linewidth]{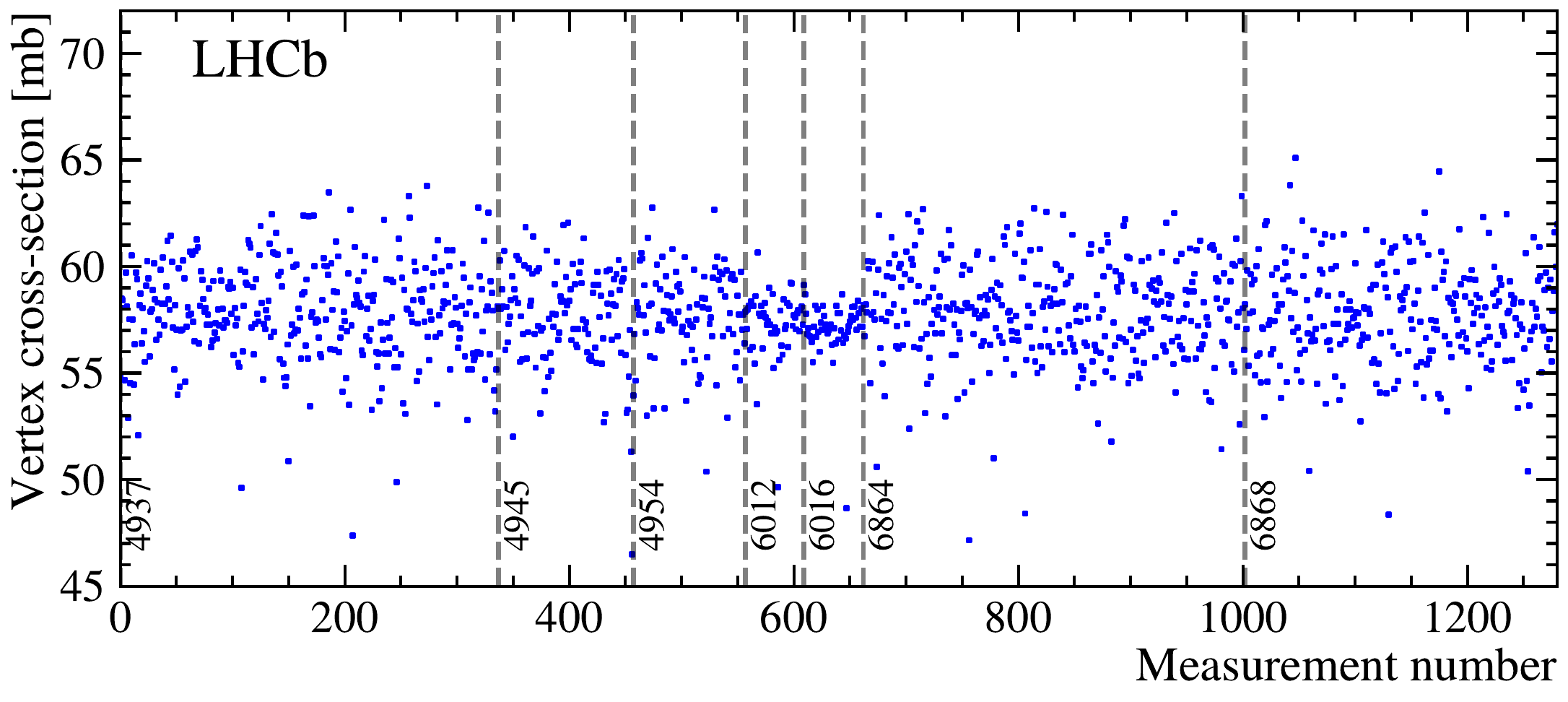}
    \caption{\small Vertex cross-section measurements using the baseline bunch profile model. Each data point corresponds to a single integration window and BCID. The data points are sorted chronologically, and then by BCID. The dashed vertical lines indicate the boundaries between the different fills included in this measurement. }
    \label{fig:xsec-permeas}
\end{figure}

\begin{figure}[htp]
    \centering
    \begin{subfigure}{0.46\linewidth}
        \centering
        \includegraphics[width=\linewidth]{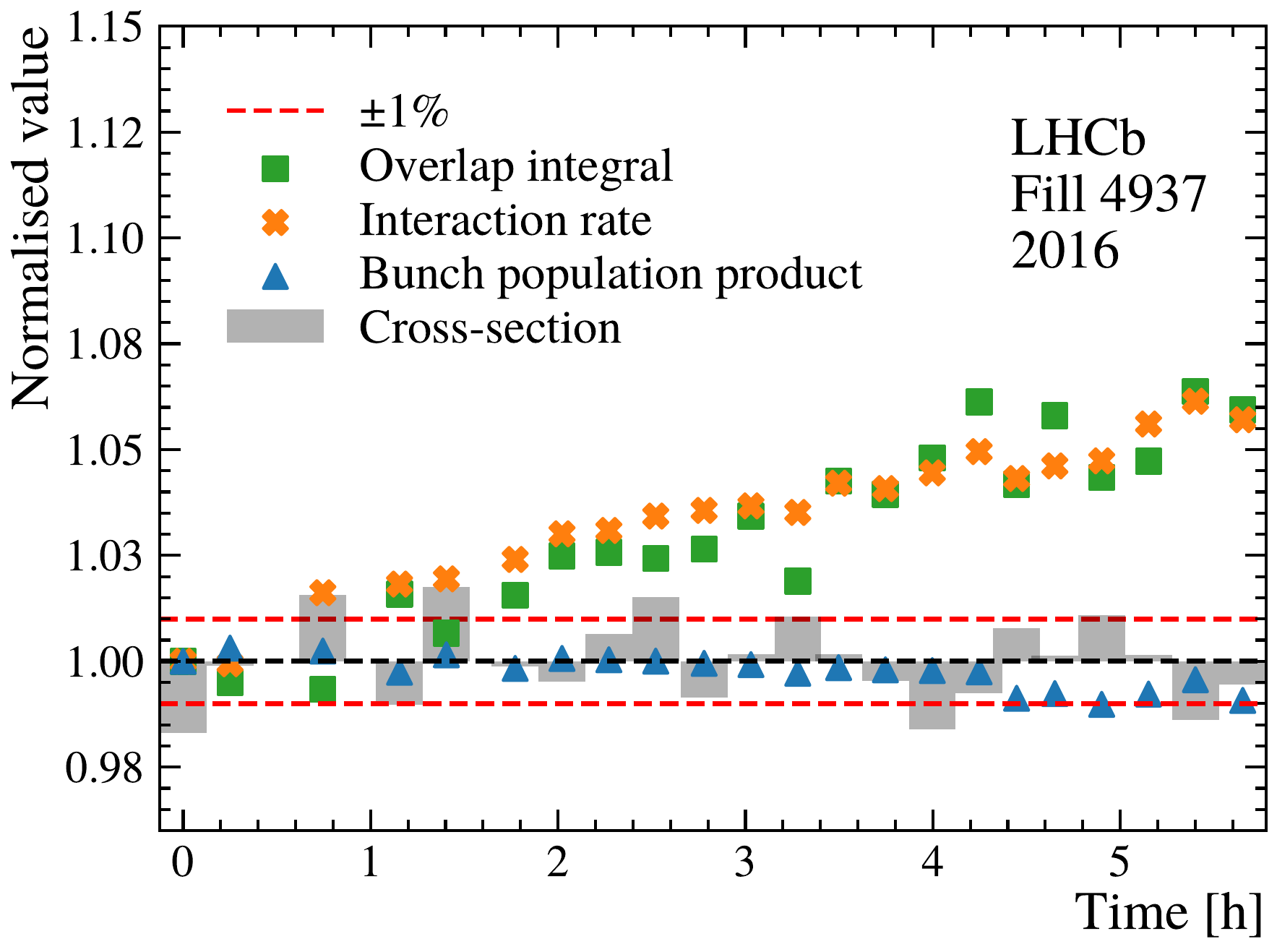}
    \end{subfigure}
    \begin{subfigure}{0.46\linewidth}
        \centering
        \includegraphics[width=\linewidth]{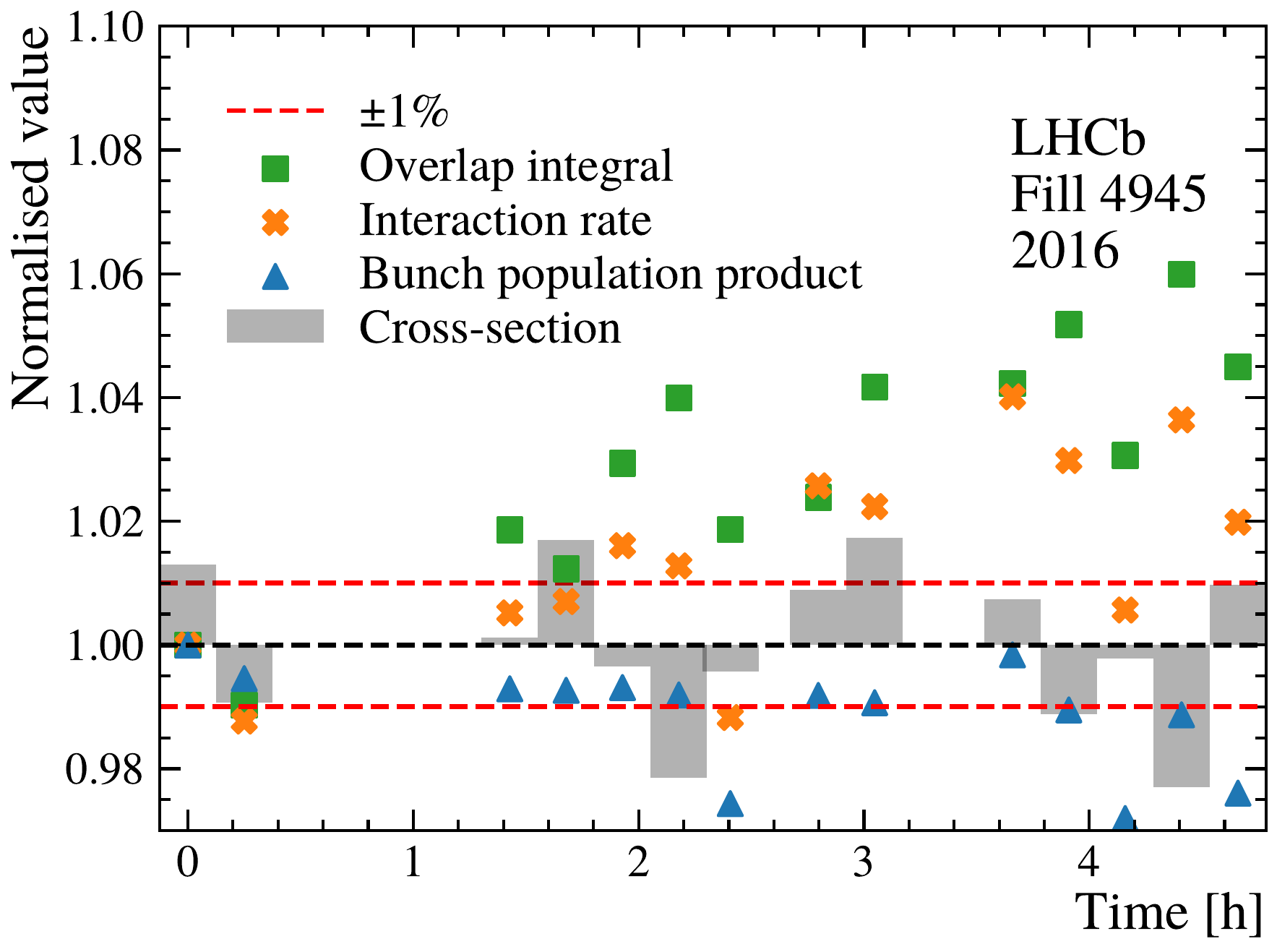}
    \end{subfigure}
    
    \begin{subfigure}{0.46\linewidth}
        \centering
        \includegraphics[width=\linewidth]{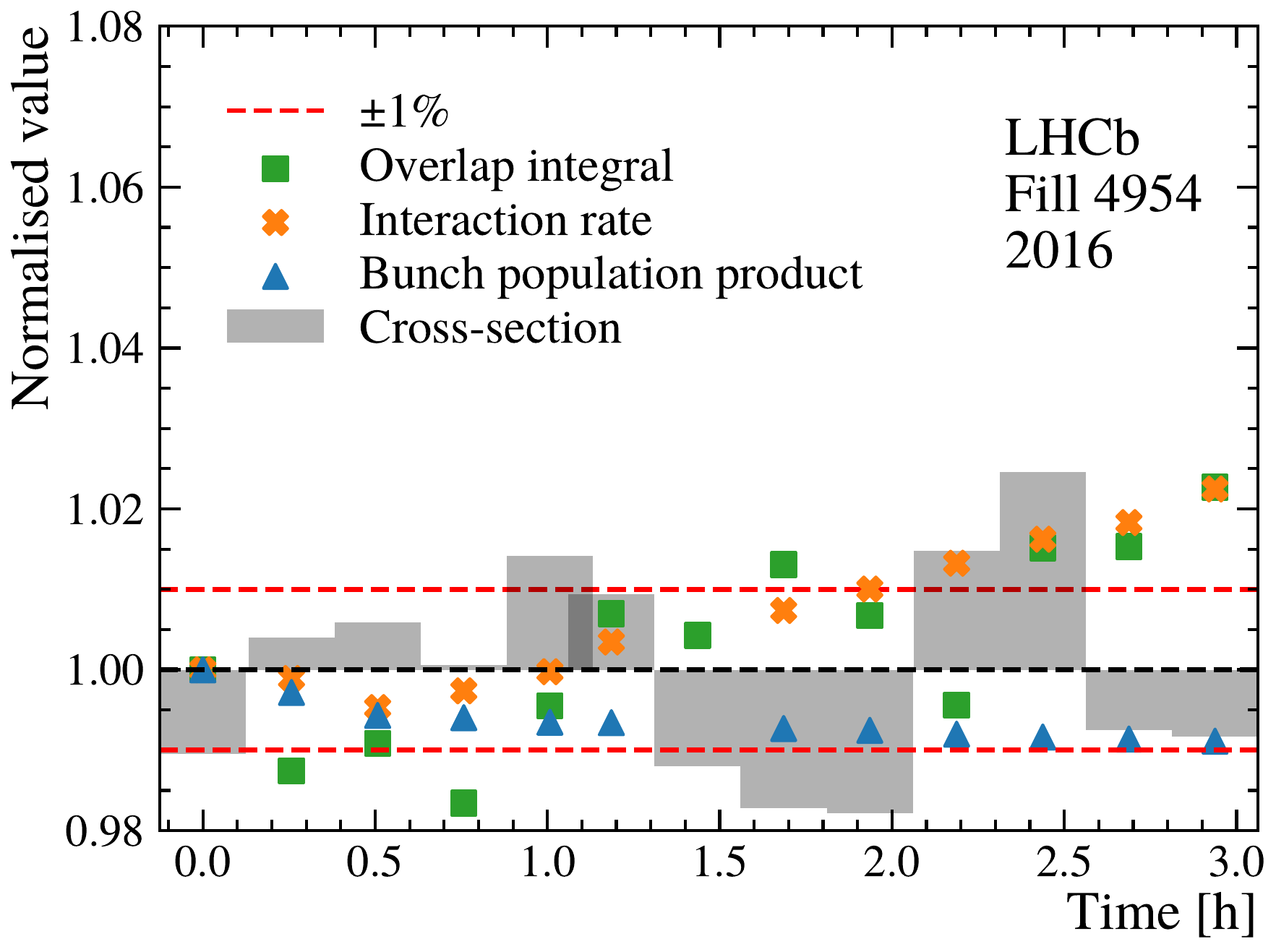}
    \end{subfigure}
    
    \begin{subfigure}{0.46\linewidth}
        \centering
        \includegraphics[width=\linewidth]{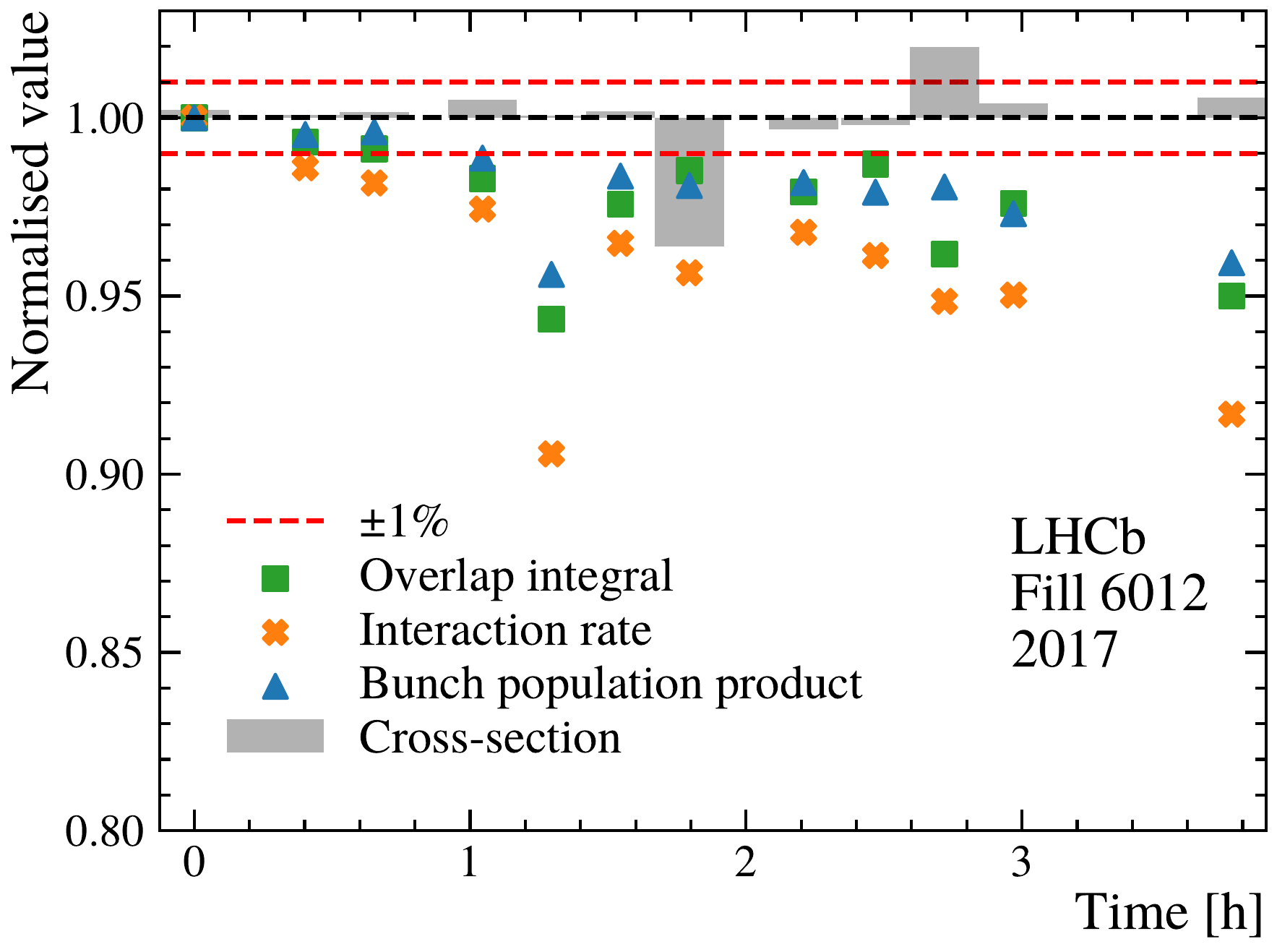}
    \end{subfigure}
    \begin{subfigure}{0.46\linewidth}
        \centering
        \includegraphics[width=\linewidth]{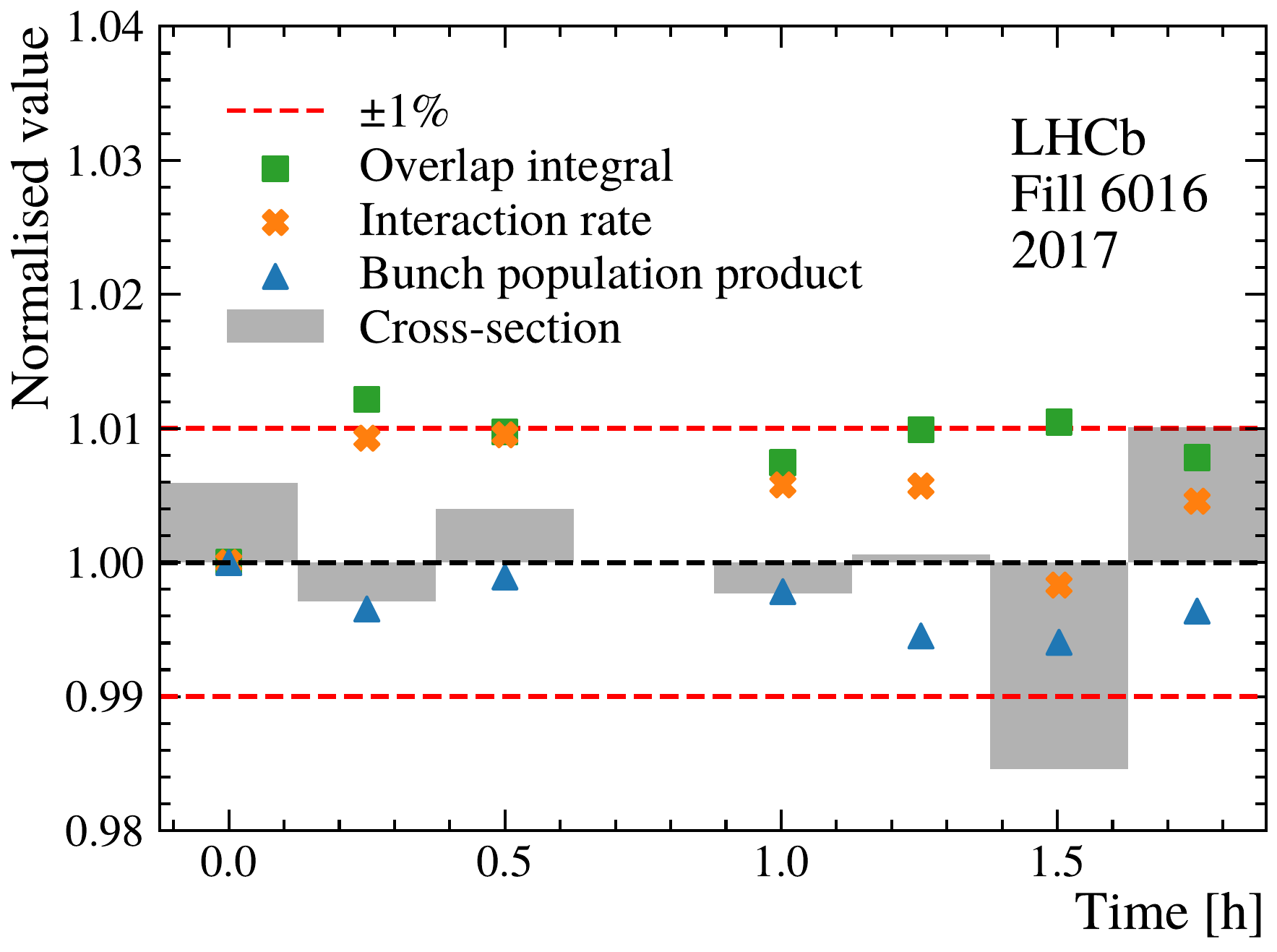}
    \end{subfigure}
    \begin{subfigure}{0.46\linewidth}
        \centering
        \includegraphics[width=\linewidth]{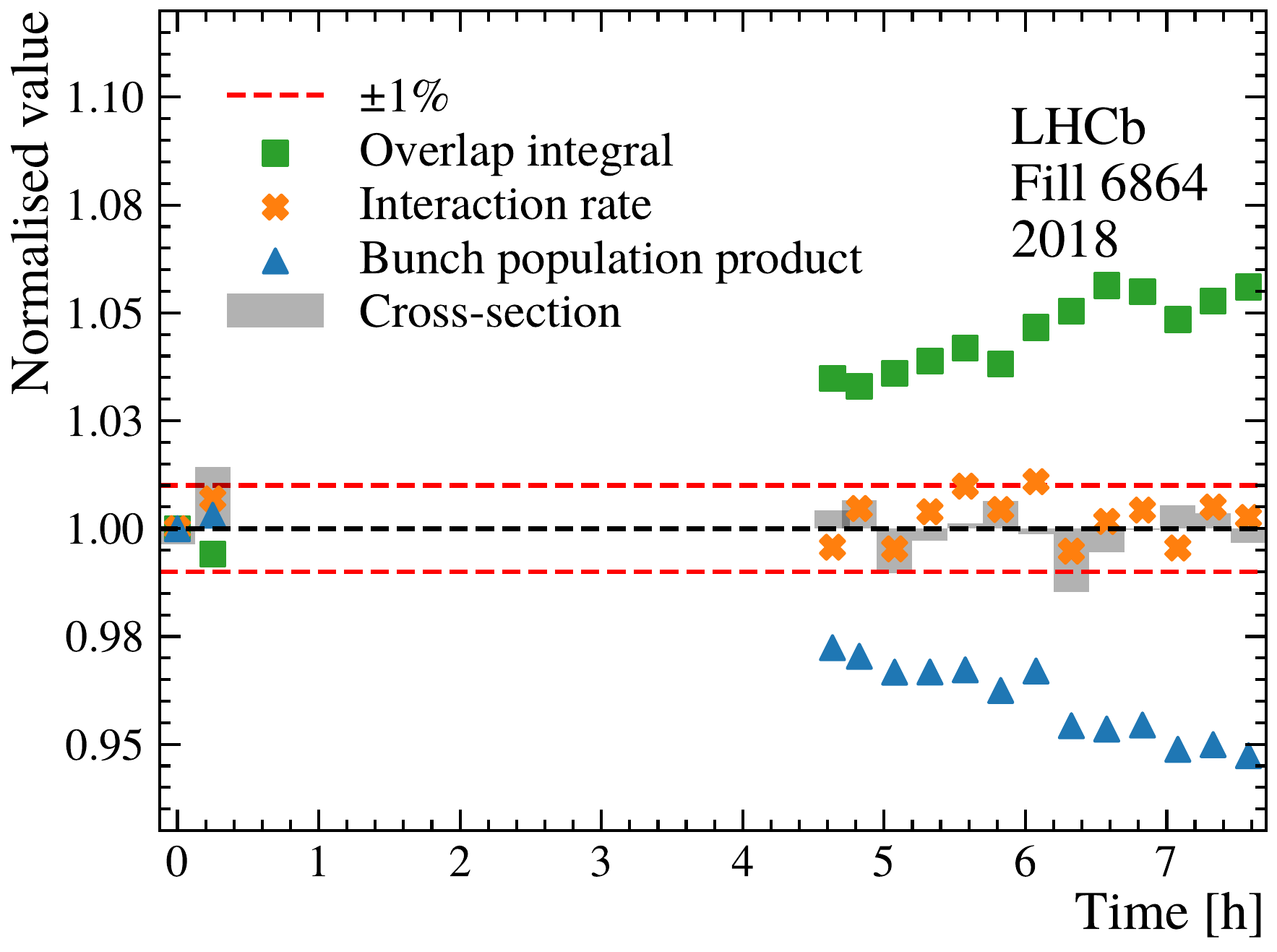}
    \end{subfigure}
    \begin{subfigure}{0.46\linewidth}
        \centering
        \includegraphics[width=\linewidth]{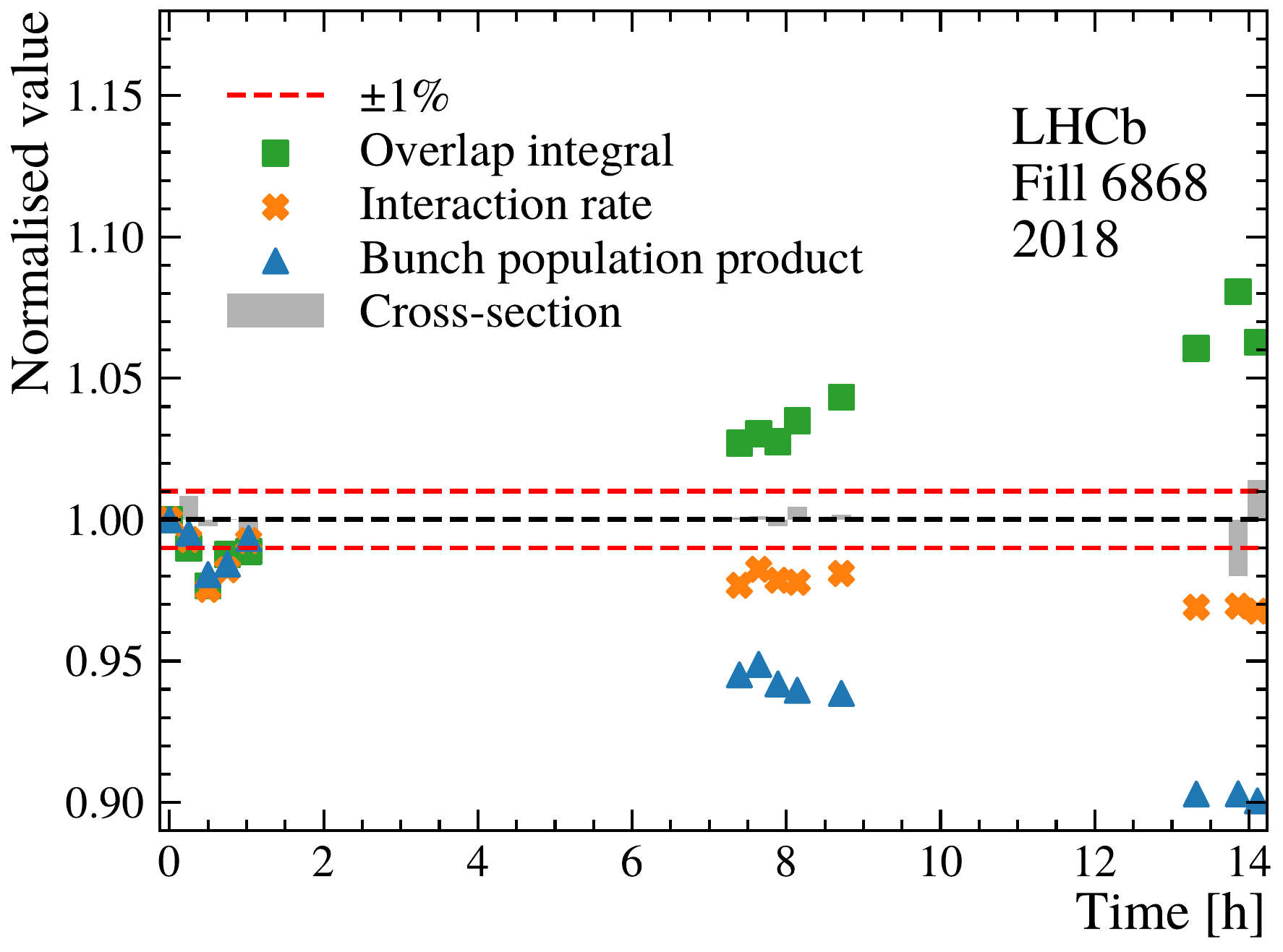}
    \end{subfigure}
    \caption{\small Time evolution of quantities entering the cross-section calculation for each fill, averaged over all BCIDs. The cross-section values are normalised to their mean value; all other quantities are normalised to their first value.}
    \label{fig:evolution}
\end{figure}

A total cross-section for each BCID is obtained from a weighted average of the individual integration periods according to their duration. Integration periods where the p-value of the fit to either of the transverse bunch profiles is less than~1\% are removed; this requirement removes approximately~2\% of all measurements. The measured cross-section for each fill is obtained as the average of the measurements for each BCID. The final measured cross-section is obtained from a fit to the per-fill cross-section using the standard statistical uncertainties only. Considering the statistical uncertainty only, the cross-section of the Vertex counter is measured to be $\sigma_\mathrm{Vertex} = 57.61 \pm 0.06 \mbarn$. Figure~\ref{fig:xsec-perfill} shows the measured value of the cross-section in each calibration fill. The results for each fill are not statistically compatible with each other. A systematic uncertainty, described later in this section, is assigned as a result of this disagreement. Figure~\ref{fig:xsec-permeas} shows each individual cross-section measurement, and Fig.~\ref{fig:evolution} shows the evolution of the BCID-averaged values of all quantities entering into the cross-section calculation throughout each of the calibration fills. The individual fill measurements are not statistically compatible, and a systematic uncertainty to account for this is detailed below. Good stability of the measured cross-section is observed throughout each fill.

A summary of the uncertainties affecting this measurement is presented in Table~\ref{tab:systematics}. A detailed explanation of each source of systematic uncertainty is provided in the following.

\begin{table}[tp]
    \centering
    \caption{\small Summary of systematic uncertainties on the absolute BGI calibration for the Vertex cross-section.}
    \label{tab:systematics}
    \begin{tabular}{cc}
        \hline\hline
        Source & Uncertainty [mb]\\
         \hline
         Vertex resolution bias & 0.29  \\
         Fill-by-fill disagreement & 0.26 \\
         Transverse profile fit model & 0.19\\
         Beam-beam effect & 0.15\\
         DCCT uncertainty & 0.12\\
         $z_\text{RF}$ approximations & 0.12\\
         Beam drift & 0.10\\
         Longitudinal efficiency & 0.07\\
         LHC frame & 0.07\\
         Equal gradients frame & 0.06\\
         Detector alignment & 0.05\\
         Counter background subtraction & 0.05\\
         Counter $z$-efficiency & 0.04\\
         Satellite charge subtraction & 0.03\\
         VELO transverse scale & 0.03\\
         Ghost charge fraction & 0.02\\
         SMOG pressure gradient & 0.02\\
         FBCT nonlinearity & 0.02 \\
         FBCT/BPTX difference & 0.01 \\
        \hline
         Total systematic uncertainty & 0.52 \\
         \hline\hline
    \end{tabular}
\end{table}

A systematic uncertainty arising from potential nonlinearities in the measurements provided by the FBCTs is determined by the change in the measured cross-section, 0.02\mbarn, when applying the prescription from Ref.~\cite{Anders:1427726}. The shift in the cross-section from the baseline measurement when using the BPTX measurements of the bunch population fractions, 0.01\mbarn, is assigned as a systematic uncertainty. The difference in the measured cross-section when all ghost charge measurements are shifted upwards and downwards by their uncertainty, 0.02\mbarn, is assigned as a systematic uncertainty. The total shift in the cross-section when neglecting the satellite charge correction, 0.03\mbarn, is assigned as a systematic uncertainty. A systematic uncertainty of 0.12\mbarn, corresponding to a relative precision of 0.20\%, on the measurements from the DCCTs is assigned according to the prescription from Ref.~\cite{Barschel:1425904}.\footnote{The DCCTs were read out in \lhc Run 1 using a 12-bit analogue-to-digital converter (ADC), which has since been replaced with an improved 24-bit ADC. To avoid a large overestimation of uncertainties dominated by the least significant bit, the uncertainties for fills 6864 and 6868 are calculated assuming DCCT range 3, while the analysed data was collected using DCCT range 2 (see Table~2 of Ref.~\cite{Barschel:1425904}). This is consistent with the approach taken in Ref.~\cite{ATLAS_vdm}.}

Potential systematic effects affecting the interaction rate background subtraction are assessed by comparing the baseline Vertex counter with an alternative counter. The alternative counter, labelled PV3D, counts the number of reconstructed PVs in each event without any restriction on reconstructed position. The PV3D counter is more sensitive to beam-gas background due to its increased acceptance outside the luminous region. The largest deviation from unity of the ratio between the two counters averaged across any calibration fill, 0.05\mbarn, is taken as a systematic uncertainty. An additional systematic uncertainty is assigned based on the difference in the final cross-section when applying the simulation- and calorimeter-based efficiency corrections to the relative luminosity, 0.04\mbarn.

The detector misalignment was found to induce a systematic uncertainty on the measurement of the crossing angle in a previous measurement~\cite{LHCb-PAPER-2014-047}. The effect of a biased crossing angle on the overlap integral, and thus the cross-section, is approximately quantified by
\begin{equation}
    \left|\frac{\delta\mathcal{O}}{\mathcal{O}}\right| \approx \left|\frac{\phi_x \lambda_z^2}{\Sigma_x^2} \right| \delta\phi_x,
\end{equation}
where any offset in the transverse direction and crossing angle in the $y$-direction have been neglected. The uncertainty on the half crossing angle, $\delta\phi_x$, is taken as the larger of standard deviation of the measurements for each run in the $x$- and $y$-directions. The other parameters are taken as the average across all measurements. A systematic uncertainty of 0.05\mbarn is assigned.

A systematic uncertainty on the transverse fit model is assessed from the shift observed in the Vertex cross-section when a different bunch profile model is assumed. Two additional models are considered. The first alternative model removes the two cross terms in Eq.~\ref{eq:baseline-model} where the signs of $\delta_x$ and $\delta_y$ are opposite, while still guaranteeing symmetry in the transverse directions. This modification allows a rotation of the transverse axes of the bunch shapes with respect to the \lhcb reference frame. The second alternative model has the same general structure, but removes the constraints that guarantee the shape to be symmetrical, introducing additional degrees of freedom to the shape. This is achieved by allowing all three Gaussian components in each transverse axis to have independent mean parameters. Shifts of $-0.06\mbarn$ and $+0.13\mbarn$, respectively, are observed, the latter of which is assigned as a systematic uncertainty associated with the choice of transverse bunch profile model.

The measurement of the longitudinal bunch profiles only enters into the calculation of the overlap integral through the presence of either a crossing angle or offset between the beams. The effect of mismodelling in the longitudinal direction on the overlap integral measurement is therefore suppressed with respect to mismodelling in the transverse directions. A systematic uncertainty related to the longitudinal efficiency correction is assigned by repeating the longitudinal fits without any efficiency correction. The resulting shift in the measured cross-section, 0.07\mbarn, is taken as a systematic uncertainty

It is assumed here that the bunch profiles are stable throughout each integration period. Any movement of the beams will artificially increase the measured bunch widths. To evaluate the size of the effect, the mean position of each beam in each axis is measured in 30\sec time intervals throughout each data-taking run, integrated across all filled BCIDs. Each run lasts up to 1\hr, corresponding to four full integration periods. The average drift across each beam, axis and run is found to be 4.1\mum. Assuming a beam width of~100\mum results in an estimated relative bias of 0.09\% on the measured bunch widths. This effect is added linearly for each of the two transverse axes, as it is fully correlated. The resulting systematic uncertainty on the final cross-section is therefore taken to be~0.10\mbarn. As an additional check, the integration period used to perform the measurement is increased and decreased by a factor of two, resulting in shifts of $+0.09\mbarn$ and $-0.02\mbarn$, respectively. This is in agreement with the systematic uncertainty quoted above, so no additional uncertainty is assigned. In principle, any drift of the beams can also bias the measurement of the transverse offsets. This effect is estimated to bias the cross-section by less than~0.01\mbarn and is therefore neglected.

The determination of $z_\text{RF}$ from Eq.~\ref{eq:lumi-mu} relies on the assumption that the bunches have equal widths in the $x$-direction. To test this assumption, a pseudoexperiment study is performed. The effect of asymmetric bunch widths is simulated, using the luminous region parameters measured in data to constrain input values of $z_\mathrm{RF}$ to a realistic range. All other parameters are taken as input from the values measured using the data. The bias on the overlap integral, by comparing the overlap integral calculated with the known $z_\mathrm{RF}$ to that computed using Eq.~\ref{eq:lumi-mu}, is found to be at most 0.12\mbarn, which is taken as a systematic uncertainty.

The measurements presented here are all performed in the \lhcb reference frame, which is rotated with respect to the \lhc reference frame in the transverse axes by~$13~\text{mrad}$. Approximate factorisation of the bunches into $x$- and $y$-directions should be most applicable in the \lhc reference frame. While the fit models used here are designed to account for nonfactorisability, this effect is nonetheless checked by repeating the measurements of the overlap integral using the baseline selection and global resolution correction in the \lhc reference frame. A shift of $0.07~\text{mb}$ is observed in the final Vertex cross-section, which is taken as a systematic uncertainty.

Similarly, the expressions used to evaluate the overlap integral rely on the assumption that the beams collide with equal and opposite gradients in the crossing plane~\cite{Ferro-Luzzi:2012zxa}. A slight deviation from this behaviour is observed in the \lhcb frame, as can be seen in Fig.~\ref{fig:example-crossing}. To estimate the size of this effect, the overlap integral measurements using the baseline selection and global resolution correction are performed after rotations in both the $xz$- and $yz$-planes such that the two beam gradients are equal and opposite. This results in a shift of~$0.06~\text{mb}$ in the final Vertex cross-section, which is taken as a systematic uncertainty.

\begin{figure}[tbp]
    \centering
    \includegraphics[width=0.6\linewidth]{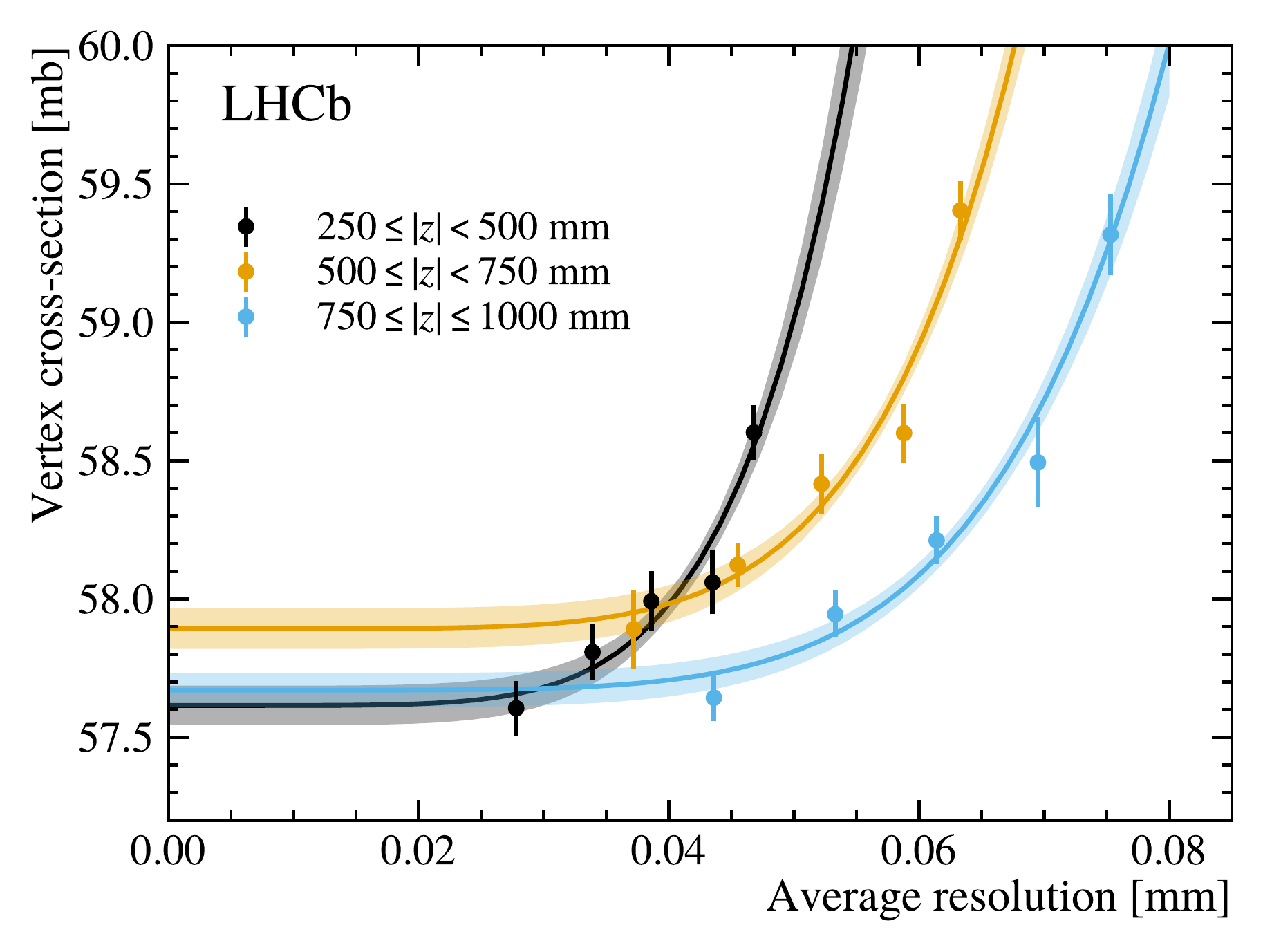}
    \caption{\small Extrapolation of the measured Vertex cross-section used to determine the systematic uncertainty associated with the beam-gas vertex resolution. The shaded bands represent the 68\% confidence interval around the central values of the fitted functions.}
    \label{fig:xsec-extrapolation}
\end{figure}

The vertex resolution depends strongly on the number of tracks in each vertex, and the longitudinal vertex position. Thus, potential biases related to the determination of the vertex resolution can be tested by studying the total cross-section as a function of selection requirements on each of these variables. The vertex resolution primarily affects the bunch profile parameters related to the transverse bunch widths. The underlying transverse bunch widths are calculated by deconvolving the resolution width as $\lambda_\mathrm{meas} = \sqrt{\lambda^2 - \tau^2}$, where $\lambda$ is the width of the distribution in data, $\lambda_\mathrm{meas}$ is the measured bunch width after deconvolution, and $\tau$ is the measured resolution width. In the limit of perfect resolution, $d\lambda_\mathrm{meas}/d\tau = 0$, and consequently $d\sigma/d\tau = 0$. Thus, assuming a polynomial expansion of the cross-section in terms of the resolution, the linear term can be assumed to be zero. It is implicitly assumed here that the measured, potentially biased, resolution tends to zero as the actual resolution tends to zero. This assumption is justified since the resolution is measured using the residual distance between a pair of fitted vertices, which will tend to zero in the limit of perfect resolution. The measurement of the cross-section is repeated with the data binned based on the number of reconstructed tracks in each beam-gas vertex. The bin with the most tracks associated to each vertex corresponds to the baseline requirement. This set of measurements is also repeated in two additional longitudinal regions, defined by $500 \leq |z| < 750$\mm and $750 \leq |z| \leq 1000$\mm. The results are shown in Fig.~\ref{fig:xsec-extrapolation}. An extrapolation is performed by assuming the cross-section can be parameterised as a function of the average resolution across both beams and axes with the form $\sigma_\mathrm{Vertex} = \sigma_0 + \sigma_1 \tau^a$, where $\tau$ is the measured resolution, and $\sigma_{0,1}$ are floated parameters. Empirically, the value $a=6$ is found to represent the data reasonably well. The largest difference between the baseline measurement and any of the extrapolated values is 0.28\mbarn. The minimum-maximum spread of the extrapolated values and the tightest track selection in each $z$-region are 0.28\mbarn and 0.29\mbarn, respectively. A systematic uncertainty of 0.29\mbarn is assigned based on potential biases in the determination of the beam-gas vertex resolution.

The effect of electromagnetic interactions between protons from opposite colliding bunch pairs contributes one of the largest systematic uncertainties in van der Meer luminosity calibration measurements~\cite{LHCb-PAPER-2014-047}. Changes in the beam-beam force due to movements of the beams at any interaction point on the \lhc ring can dynamically modify the bunch profiles and therefore bias this measurement. To evaluate the importance of this effect, the cross-section is measured with BCIDs split into two categories depending on whether they collide only at the \lhcb IR, or also in other experiments. Cross-sections for the Vertex counter of $57.57 \pm 0.10 \mbarn$ and $57.79 \pm 0.08 \mbarn$ are obtained for each category, respectively, where only statistical uncertainties are considered. Since these values are statistically compatible, no systematic uncertainty is assigned.

An additional bias is possible from the beam-beam interactions, due to possible, static, induced non-Gaussian shapes of the bunch profiles. To assess the related systematic uncertainty, sets of pseudoexperiments are performed. In each pseudoexperiment, a data sample representing a bunch from each beam is generated and the beam-beam force is simulated over many turns of the \lhc ring~\cite{BxB}. The simulations are performed both including interactions at other IRs on the \lhc ring and with only collisions at the \lhcb IR simulated. The baseline pseudoexperiments use a set of input parameters corresponding to the average conditions across all calibration fills. The pseudoexperiments are repeated varying the input parameters either within their uncertainties, or to their extremum values across all measurements, as appropriate. The bias is assessed by determining the overlap integral via fitting the output distributions from the simulation with the baseline transverse bunch profile model. The crossing angle is neglected in the simulations, which removes any dependence on the longitudinal bunch profiles. The resulting bias is calculated as the difference from unity of the ratio of the overlap integral determined by fitting the output of the simulation, with a numerical calculation of the overlap integral from the simulation. Averaging across all pseudoexperiments, a systematic uncertainty of 0.15\mbarn is assigned.

As visible in Fig.~\ref{fig:xsec-perfill}, the measurements of the cross-section are not statistically compatible among all of the calibration fills. This is believed to be due to mismodelling of the bunch profiles. Irrespective of the source of this effect, an additional systematic uncertainty is assigned based upon the observed spread of the measurements. The spread is considered when the measurements are split by fill, by data-taking run and by BCID; in the latter two cases, the individual measurements are found to be statistically compatible. The systematic uncertainty is therefore taken as the standard deviation of the measurements from the different calibration fills, 0.26\mbarn.

An additional pair of systematic uncertainties of 0.02\mbarn and 0.03\mbarn, related to a possible pressure gradient of the gas within the beam volume and the knowledge of the absolute scale of the \velo, respectively, are assigned. 

\section{Conclusions}
An absolute luminosity calibration of the \lhcb Vertex counter in \proton\proton collisions at ${\sqrt{s} = 13\tev}$ during Run 2 of the \lhc has been presented. The cross-section of the Vertex counter is measured to be
\begin{equation*}
    \sigma_\mathrm{Vertex} = 57.61 \pm 0.06 \stat \pm 0.52 \syst \mbarn.
\end{equation*}
The total relative precision of this measurement is 0.91\%. The dominant systematic uncertainties arise from imperfect knowledge of the beam-gas vertex resolution, the observed fill-by-fill spread of the measured cross-section, and the choice of transverse fit profile model. The largest source of uncertainty which would be correlated with a measurement of the Vertex cross-section using the vdM approach, related to the bunch current normalisation based on measurements from the DCCTs, is subdominant. To date, measurements of absolute cross-sections at \lhcb using Run~2 data have used a preliminary measurement of the absolute luminosity with an uncertainty of 2\%. Already, after propagation to physics data-taking and including relevant additional systematic uncertainties, this result would serve to reduce the systematic uncertainty on these measurements associated with the luminosity by around a factor of two. Further improvement is expected when combining this result with a future measurement which utilises the vdM approach.

%% file: LHCb/acknowledgements.tex
\section*{Acknowledgements}
%
%
\noindent We express our gratitude to our colleagues in the CERN
accelerator departments for the excellent performance of the LHC. We
thank the technical and administrative staff at the LHCb
institutes.
We acknowledge support from CERN and from the national agencies:
ARC (Australia);
CAPES, CNPq, FAPERJ and FINEP (Brazil); 
MOST and NSFC (China); 
CNRS/IN2P3 and CEA (France);  
BMFTR, DFG and MPG (Germany);
NKFIH (Hungary);              
INFN (Italy); 
NWO (Netherlands); 
MNiSW and NCN (Poland); 
MEC/IFA (Romania); 
MICIU and AEI (Spain);
SNSF and SER (Switzerland); 
NASU (Ukraine); 
STFC (United Kingdom); 
DOE NP and NSF (USA).
We acknowledge the computing resources that are provided by ARDC (Australia), 
CBPF (Brazil),
CERN, 
IHEP and LZU (China),
IN2P3 (France), 
KIT and DESY (Germany), 
INFN (Italy), 
SURF (Netherlands),
Polish WLCG (Poland),
IFIN-HH (Romania), 
PIC (Spain), CSCS (Switzerland), 
GridPP (United Kingdom),
and NSF (USA).  
We are indebted to the communities behind the multiple open-source
software packages on which we depend.
Individual groups or members have received support from
RTP (Australia), 
FWO Odysseus grant G0ASD25N (Belgium), 
Key Research Program of Frontier Sciences of CAS, CAS PIFI, CAS CCEPP (China); 
Minciencias (Colombia);
EPLANET, Marie Sk\l{}odowska-Curie Actions, ERC and NextGenerationEU (European Union);
A*MIDEX, ANR, IPhU and Labex P2IO, and R\'{e}gion Auvergne-Rh\^{o}ne-Alpes (France);
Alexander-von-Humboldt Foundation (Germany);
ICSC (Italy); 
Severo Ochoa and Mar\'ia de Maeztu Units of Excellence, GVA, XuntaGal, GENCAT, InTalent-Inditex and Prog.~Atracci\'on Talento CM (Spain);
the Leverhulme Trust, the Royal Society and UKRI (United Kingdom).

%% file: Authorship_LHCb-PAPER-2026-025.tex
\centerline
{\large\bf LHCb collaboration}
\begin
{flushleft}
\small
R.~Aaij$^{39}$\lhcborcid{0000-0003-0533-1952},
M.~Abdelfatah$^{71}$,
A.S.W.~Abdelmotteleb$^{59}$\lhcborcid{0000-0001-7905-0542},
C.~Abellan~Beteta$^{53}$\lhcborcid{0009-0009-0869-6798},
F.~Abudin\'en$^{61}$\lhcborcid{0000-0002-6737-3528},
T.~Ackernley$^{63}$\lhcborcid{0000-0002-5951-3498},
A.A.~Adefisoye$^{71}$\lhcborcid{0000-0003-2448-1550},
B.~Adeva$^{49}$\lhcborcid{0000-0001-9756-3712},
M.~Adinolfi$^{57}$\lhcborcid{0000-0002-1326-1264},
P.~Adlarson$^{87,44}$\lhcborcid{0000-0001-6280-3851}
% ...

{\footnotesize \it

% ...
$^{39}$Nikhef National Institute for Subatomic Physics, Amsterdam, Netherlands\\
% ...
$^{44}$National Center for Nuclear Research (NCBJ), Warsaw, Poland\\
% ...
$^{49}$Instituto Galego de F{\'\i}sica de Altas Enerx{\'\i}as (IGFAE), Universidade de Santiago de Compostela, Santiago de Compostela, Spain\\
% ...
$^{53}$Physik-Institut, Universit{\"a}t Z{\"u}rich, Z{\"u}rich, Switzerland\\
% ...
$^{57}$H.H. Wills Physics Laboratory, University of Bristol, Bristol, United Kingdom\\
% ...
$^{59}$Department of Physics, University of Warwick, Coventry, United Kingdom\\
% ...
$^{61}$School of Physics and Astronomy, University of Edinburgh, Edinburgh, United Kingdom\\
$^{62}$School of Physics and Astronomy, University of Glasgow, Glasgow, United Kingdom\\
$^{63}$Oliver Lodge Laboratory, University of Liverpool, Liverpool, United Kingdom\\
% ...
$^{71}$Syracuse University, Syracuse, NY, United States\\
% ...
$^{87}$Department of Physics and Astronomy, Uppsala University, Uppsala, Sweden, associated to $^{62}$\\
% ...
}
\end{flushleft}